\documentclass[fleqn,usenatbib]{mnras}

\usepackage{epsfig}
\usepackage{times}
\usepackage{amsmath}
\usepackage[british]{babel}             
\usepackage[british]{babel}             
\usepackage{newtxtext}                  
\usepackage[slantedGreek]{newtxmath}    
\usepackage{threeparttable}
\usepackage{subfig}
\usepackage{threeparttable}

\let\la=\lesssim 
\usepackage[T1]{fontenc}                
\usepackage{graphicx}                   
\usepackage{subfig}
\usepackage{hyperref}

\newif\ifAMStwofonts

\newcommand{\target}{PSR\,J1023+0038}

\newcommand{\kms}{\,km\,s$^{-1}$}

\newcommand{\Msun}{\,$\rm M_{\sun}$}
\newcommand{\Rsun}{\,$\rm R_{\sun}$}

\newcommand{\teff}{$T_{\rm eff}$}

\title[Chemical abundance analysis of PSR\,J1023+0038]
{The peculiar chemical abundance  of the transitional millisecond pulsar 
PSR\,J1023+0038 - Li enhancement}

\author[T. Shahbaz et al. ]
       {T. Shahbaz,$^{1,2}$\thanks{E-mail: tsh@iac.es}
        J. I. Gonz{\'a}lez-Hern{\'a}ndez,$^{1,2}$ 
        R. P. Breton,$^3$  
        M. R. Kennedy,$^{4,3}$  
        D. Mata S{\'a}nchez$^{1,2}$  and 
 \newauthor
        M. Linares$^{5,6}$ 
\\  
$^{ \it 1}$Instituto de Astrof\'\i{}sica de Canarias (IAC), E-38200 La Laguna, 
Tenerife, Spain \\
$^{ \it 2}$Departamento de  Astrof\'\i{}sica, Universidad de La Laguna (ULL), 
E-38206 La Laguna, Tenerife, Spain \\
$^{ \it 3}$Jodrell Bank Centre for Astrophysics, School of Physics and 
Astronomy, The University of Manchester, Manchester M13 9PL, UK  \\
$^{ \it 4}$Department of Physics, University College Cork, Cork, Ireland \\
$^{ \it 5}$Institutt for Fysikk, Norwegian University of Science and Technology, Trondheim, Norway\\
$^{ \it 6}$Departament de F{\'i}sica, EEBE, Universitat Polit{\`e}cnica de Catalunya, Av. Eduard Maristany 16, E-08019 Barcelona, Spain
}

\date{Accepted XXX. Received YYY; in original form ZZZ}

\pubyear{2021}

\begin{document}
\label{firstpage}
\pagerange{\pageref{firstpage}--\pageref{lastpage}}
\maketitle

\begin{abstract} 
\noindent
Using high-resolution optical spectroscopy we determine the chemical abundance of the secondary star in the binary millisecond pulsar \target. We measure a metallicity of [Fe/H] = 0.48 $\pm$ 0.04 which is higher than the Solar value and in general find that the element abundances are different compared to the secondary stars in X-ray binaries and stars in the solar neighbourhood of similar Fe content. Our results suggest that the pulsar was formed in a supernova explosion. We find that supernova models, where matter that has been processed in the supernova is captured by the secondary star leading to abundance anomalies, qualitatively agree  with the observations. We  measure Li abundance of A(Li) = 3.66 $\pm$ 0.20, which is anomalously high compared to the Li abundance of stars with the same effective temperature, irrespective of the age of the system. Furthermore, the Li abundance in \target\ is higher than the Cosmic value and what is observed in young Population I stars  and so provides unambiguous evidence for fresh Li production. The most likely explanation is the interaction of high energy gamma-rays or relativistic protons from the pulsar wind or intrabinary shock with the CNO nuclei in the secondary star's atmosphere via spallation which leads to substantial Li enrichment in the secondary star's atmosphere.
\end{abstract}


\section{Introduction}
\label{sec:intro}

It is generally accepted that binary millisecond pulsars (MSPs) are  produced via a recycling scenario which follows the evolution of low-mass X-ray binaries (LMXBs) containing a slowly rotating neutron star. In these systems  a phase of heavy mass accretion from an evolving, non-degenerate, Roche Lobe filling secondary star eventually spun up the pulsar to millisecond spin periods \citep{Alpar82,Radhakrishnan82,Bhattacharya91}. MSPs can also be formed by accretion-induced collapse of massive white dwarfs \citep{Smedley15}. By accreting material from the secondary star, the white dwarf increases its mass above the Chandrasekhar mass limit and electron capture leads to collapse of the white dwarf, resulting in a low-field, rapidly rotating neutron star \citep{Nomoto91,Ferrario07}. Indeed, \citet{Hurley10} investigated formation rates of accretion-induced collapse neutron stars using population synthesis methods and found that accretion-induced collapse systems provide a complementary important contribution to the formation pathway to MSPs than recycling. 
 
Binary MSPs are divided into classes according to the nature of their secondary star, which can be either a degenerate or a non-degenerate object.  The `redback' class have relatively massive $>$0.2\Msun, non-degenerate secondary stars, whereas the `black widow' class contain semi-degenerate $<$0.1\Msun\ low-mass secondary stars which have suffered significant mass-loss and ablation from the pulsar wind \citep{Roberts13}. Finally there are MSPs with degenerate, extremely low-mass ($<$0.3\Msun) helium-core white dwarf companions \citep{Brown13}. Currently there is a population of 22 redbacks \citep{LK21} of which 3 (at the time of writing) are transitional MSPs, switching between distinct states of accretion-powered and rotation-powered emission on time-scales of years. Indeed, the transitional MSPs have provided support for the recycling scenario \citep{Archibald09,Papitto13,Bassa14,Stappers14,Patruno14}. For a full review of the transitional MSPs see \citet{Campana18}. It is thought that irradiation feedback \citep{Benvenuto14} and thermal-viscous instability in the accretion disc \citep{Jia15} is responsible for the transitions between the two states. X-rays  absorbed by the secondary star produce a partial blockage of the energy arising from the star's interior leading to cyclic mass-transfer episodes between the X-ray binary and radio pulsar states \citep{Bunning04,Benvenuto14}. 

In the rotation-powered pulsar-state (PS) regular radio eclipses and pulsations are observed \citep{Archibald09,Archibald13}. The stripping of mass from the companion as a result of bombardment by the energetic wind of the MSP causes the companion star to bloat and is the favoured mass-loss mechanism \citep{Ruderman89}. This mass-loss gives rise to long eclipses at superior conjunction, often covering most of the orbit \citep{Polzin18, Polzin20}. The optical light curves show an asymmetric single-humped modulation, due to the combined effects of the  nearly Roche lobe filling tidally-locked secondary star's ellipsoidal modulation and the high-energy emission from the pulsar wind heating of  the inner face of the secondary star \citep{Breton13,Schroeder14,Strader19}. In multiple systems, the asymmetric light curves suggests that the heating does not come directly from the isotropic pulsar wind, but is due to non-thermal X-ray emission produced by the intrabinary shock between the pulsar and secondary star's 
wind \citep{Romani16}, or gets redistributed around the star 
\citep{Kandel19, Voisin20}.

\begin{figure}
\centering
\includegraphics[width=1.0\linewidth,angle=0]{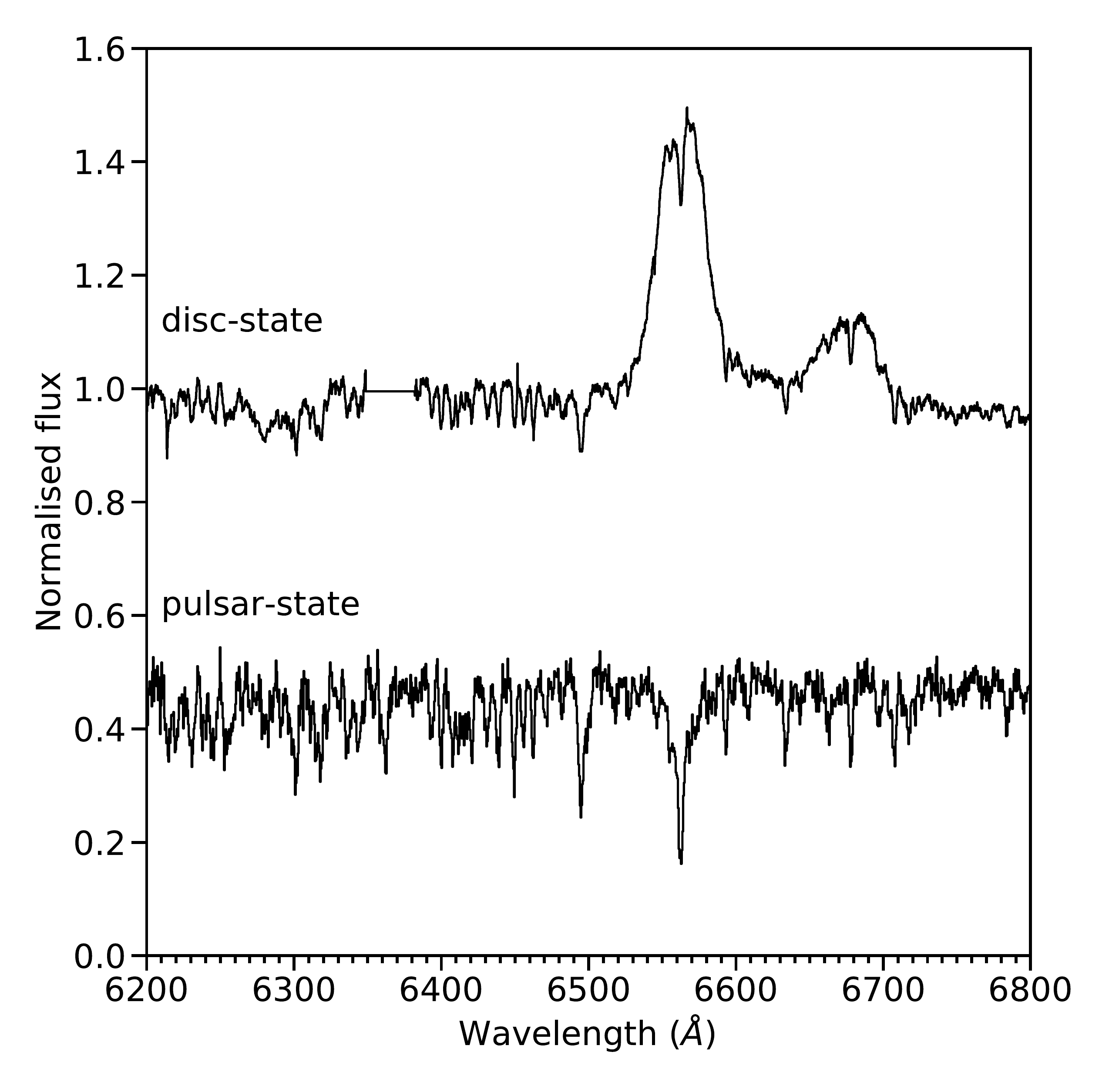}
\caption{
The 2009 pulsar-state spectrum (bottom) and 2016 disc-state spectrum (bottom), shifted vertically for clarity. 
The disc-state spectrum contains a gap between 6350--6400\AA\ due to bad columns in the CCD.
Only part of the spectra are shown. 
} 
\label{fig:full_spectrum}
\end{figure}

In the accretion-powered disc-state (DS), no radio pulsations are detected and the system is much brighter at optical and X-ray wavelengths due to the presence of an accretion disc. X-ray pulsations at the neutron star spin period are observed \citep{Archibald09}. The secondary star is close to fully filling its Roche lobe and the accretion disc is observed via broad emission lines. The optical light curve is similar to that in the pulsar-state suggesting that the level of irradiation/heating of the secondary star is similar \citep{Kennedy18}. The effects of heating are observed in the optical spectra \citep{Shahbaz19} but it is not clear if the source of the irradiation is the high-energy emission from the pulsar relativistic wind or the X-ray emission from inner regions of the accretion disc. Low- and high-mode flux variations \citep{Bogdanov15,Shahbaz15} and coherent optical and X-ray pulsations are observed \citep{Ambrosino17,Archibald15} as well as GeV emission  \citep{Takata14,Papitto19}. Recently, two similar models have been proposed to explain the properties observed in \target\ \citep{Veledina19,Papitto19}; the dissipative collision between a rotating striped pulsar wind at a few light cylinder radii away from the pulsar, gives rise synchrotron emission producing the optical and X-ray pulses. 

The observed properties of binary MSPs depend on their evolutionary path.  Indeed, the way in which the neutron star is formed, the mass of the donor star and the orbital separation after the birth of the neutron star, dictates the evolutionary state of the system. It is generally believed that in a binary  consisting of a massive star and a low-mass donor star, mass transfer will become dynamically unstable and the binary undergoes a phase of common envelope evolution.  During this phase the donor star spirals in towards the centre of the massive star ($>$10\Msun) ejecting most of its envelope, leaving behind a naked He-burning core.  A neutron star is subsequently born after the explosion of the helium star. If the binary is not  disrupted and if the orbital separation is small enough, the (evolved) non-degenerate donor star fills its Roche lobe and the binary undergoes a subsequent epoch of mass transfer on to the neutron star. In this phase the system is observed as an X-ray binary \citep{Tauris06}.

The donor star is expected to loose a significant amount of its outermost layers, a small fraction of which is accreted by the neutron star. In a  strongly peeled star, we observe the chemical composition that has been previously modified  by thermonuclear reactions. In material partially processed by the CNO cycle, C depletion and N enhancement is  expected \citep{Chen13}. It is also possible that the SN explosion producing the neutron star modifies the chemistry of the donor star. The chemical abundances of the donor/secondary star in black hole and neutron star X-ray binaries  has  been studied in several systems: Nova Scorpii 1994 \citep{Israelian99,J1655}, A0620-00 \citep{A0620}, Cen\,X--4 \citep{CenX4}, XTE\,J1118+480 \citep{J1118_1, J1118_2}, Cyg\,X--2 \citep{CygX2}, V404\,Cyg \citep{V404Cyg}, and V4641\,Sgr \citep{Orosz01,Sadakane06}, taking into account different scenarios of pollution from supernova (SN) or hypernova (HN) ejecta.  The chemical abundances of secondary stars show over-abundance of several $\alpha$-elements (such as O, S, Si) which has been interpreted as pollution by matter ejected during the SN \citep[see][and references within]{V404Cyg}. 

The only MSP where a spectroscopic chemical abundance analysis exists is for PSR\,J1740--5340, which lies in the globular cluster NGC\,6397 \citep{DAmico01a}. PSR\,J1740--5340 is a redback MSP which has a non-degenerate secondary star ($\sim$0.3\Msun) in a 32.5\,hr binary orbit \citep{Ferraro01,Orosz03}. Anomalous Li, Ca and C abundances are observed and a comparison with theoretical models indicates that the companion is the low-mass remnant star of a deeply peeled  0.8\Msun\ progenitor \citep{Mucciarelli13}. 
In this paper we use high-resolution spectra to derive the stellar parameters and chemical abundances of the secondary star in the binary MSP  \target. We compare the abundances with stars in the Solar neighbourhood as well as the secondary stars in X-ray binaries. We also compare the determined abundances in the context of enrichment of the secondary star from SN/HN yields as well as CNO evolution models.
Finally we investigate the viability of various Li enrichment scenarios via spallation of the CNO nuclei in the atmosphere of the secondary star. 

\begin{figure*}
\centering
\begin{tabular}{c}
\includegraphics[width=0.9\linewidth,angle=0]{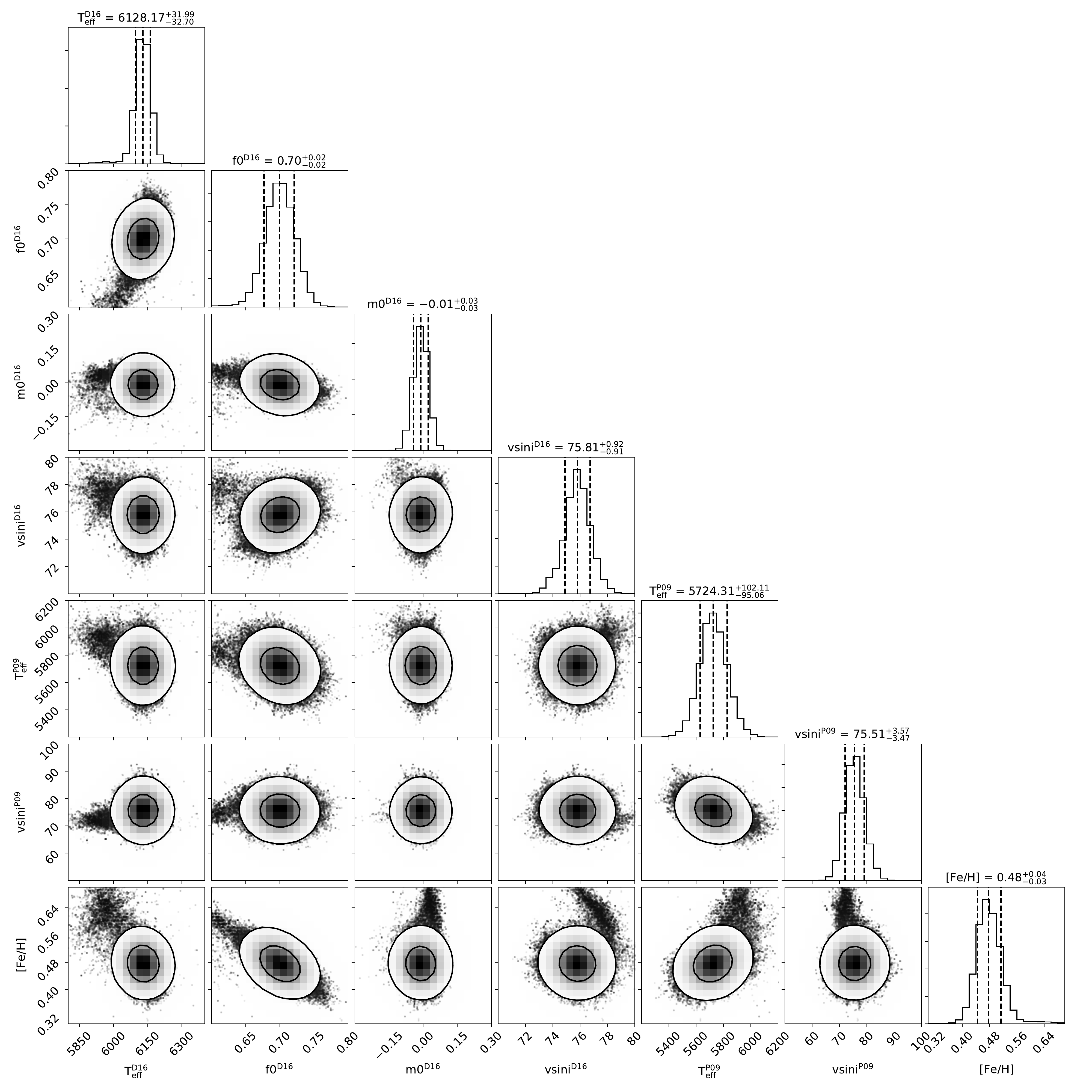}
\end{tabular}
\caption{
MCMC 2-D model parameter distributions resulting from synthetic fits to the observed spectrum of \target\ to determine the stellar parameters. The contours in the 2-D plots show the 1 and 2 sigma confidence regions, and the projected 1-D parameter distributions with the mean and standard deviation. The superscript D16 and D09 refer to the 2016 disc-state (D16) and 2009 pulsar-state (P09), respectively.
}
\label{fig:MCMC_stellar}
\end{figure*}

\begin{table}
\footnotesize
\centering
\caption{The results of the MCMC analysis to determine the stellar parameters of \target,}
\centering
\begin{tabular}{l c c}
\hline 
Parameter                    &  pulsar-state            &  disc-state             \\
                             &     2009                 &     2016                \\
\hline
 $T_{\rm eff}$ (K)           &  5724$^{+102}_{-95}$     &  6128$^{+32}_{-33}$     \\ 
 $v_{\rm rot}\,\rm sin\,{\it i}$ 
 (km\,s$^{-1}$)              &  75.5$^{+3.6}_{-3.5}$    &  75.8$^{+0.9}_{-0.9}$   \\ 
 $f_{\rm 6000}$              &  1.0                     &  0.70$^{+0.02}_{-0.02}$ \\ 
 $m_{\rm 0}$ $\times10^{-3}$ &  0.0                     & $-$0.01$^{+0.03}_{-0.03}$ \\ 
\hline
\end{tabular}
\label{table:results}
\end{table}

\begin{table}
\footnotesize
\centering
\caption{LTE element abundances determined for \target.}
\begin{tabular}{l c r r r}
\hline
  Element  & $\log\epsilon({\rm X})_{\odot}^{\rm a}$ & [X/Fe] & [X/H]$^{\rm b}$ & A(X)$^{\rm c}$  \\ 
\hline
 Li$^{\rm d}$    & 3.31$^{\rm f}$ &   0.02 $\pm$ 0.20  &   0.50 $\pm$ 0.20   &  3.66 $\pm$ 0.20 \\  
 O$^{\rm d}$     & 8.74$^{\rm e}$ &  $-$1.12 $\pm$ 0.10  &  $-$0.64 $\pm$ 0.11   &  8.10 $\pm$ 0.11 \\ 
 Al        & 6.47     &   0.87 $\pm$ 0.08  &   1.35 $\pm$ 0.09   &  7.82 $\pm$ 0.09 \\ 
 Si        & 7.55     &   0.02 $\pm$ 0.05  &   0.50 $\pm$ 0.06   &  8.05 $\pm$ 0.06 \\   
 Ca        & 6.36     &   0.14 $\pm$ 0.05  &   0.62 $\pm$ 0.06   &  6.98 $\pm$ 0.06 \\   
 Fe        & 7.50     &                    &   0.48 $\pm$ 0.04   &  7.98 $\pm$ 0.04 \\   
 Ni        & 6.25     &   0.47 $\pm$ 0.23  &   0.95 $\pm$ 0.24   &  7.20 $\pm$ 0.23 \\ 
\hline 
\end{tabular}
\begin{tablenotes}
  \item $^{\rm a}$Photospheric Solar abundances adopted from \citet{Grevesse96}.
  \item $^{\rm b}$[X/H] = log[$N$(X)/$N$(H)]-log[$N$(X)/$N$(H)]$_\odot$, where $N$(X) is the number
density of atoms.
  \item $^{\rm c}$Abundance expressed as $\rm A(X)$  = $\log$ [$N$(X)/$N$(H)] + 12.
  \item $^{\rm d}$Corrected for NLTE effects.
  \item $^{\rm e}$Solar abundance from \citet{Ecuvillon06}.
  \item $^{\rm f}$Meteoritic Solar abundance from \citet{Grevesse96}.
\end{tablenotes}
\label{table:X}
\end{table}

\section{Chemical Analysis}

The spectra analysed here have been presented in \citet{Shahbaz19} and 
so we refer the reader to it for the  data reduction details.
The pulsar-state spectrum covers spectral range 6200--6900\,\AA\ with velocity resolution of 24\kms\ whereas the disc-state spectrum covers 5300--10200\,\AA\ with a velocity resolution of 40\kms. The disc-state spectra clearly show broad Balmer and He lines emission lines related to the presence of an accretion disc. The spectral range  6200--6900\,\AA\ is common to both the pulsar- and disc-state spectra. Both data sets cover the full orbital phase range, have been Doppler-corrected to the secondary star's rest frame \citep{Shahbaz19} and carefully normalized using a low-order spline fit to the continuum level (see Fig.\,\ref{fig:full_spectrum}).

\subsection{Stellar Parameters}
\label{sec:stellar}

The chemical abundance of various elements in the secondary stars of X-ray binaries have been determined \citep[see][and references within]{V404Cyg}. We use a similar method to perform a chemical abundance analysis of the secondary star in \target\ in the pulsar- and disc-state. We  simultaneously fit the 2009 pulsar- and 2016 disc-state spectra with a synthetic model to first determine the stellar parameters and then  determine the chemical abundance of the secondary star in each state.

We determine the stellar parameters, effective temperature $T_{\rm eff}$, surface gravity $\log\,g$ and the [Fe/H] metallicity, using synthetic spectral fits to the observed spectra.  To compute the synthetic spectra we use the 2014 version of the 1D local thermodynamic equilibrium (LTE) code \textsc{moog} version 2013 \footnote{\url{https://www.as.utexas.edu/~chris/moog.html}} \citep{Sneden12}. We use the atomic line data from the Vienna Atomic Line Database VALD-2;\footnote{\url{http://www.astro.univie.ac.at/~vald}} \citep{Kupka00} 
and a grid of local thermodynamic equilibrium (LTE) model atmospheres provided by \citet{Kurucz93}. The model grid spans $T_{\rm eff}$ = 5000--7000\,K in steps of 100\,K, $\log\,g$ = 3.0--5.0 $\rm cm\,s^{-2}$ in steps of 0.1\,dex and [Fe/H] abundances between $-$0.5 and 1.0 in steps of 0.1\,dex. The micro-turbulence is computed using the expression which depends on effective temperature and surface gravity \citep{Adibekyan12}. Subsequently, for a given model atmosphere we compute the synthetic spectra using \textsc{moog}.

It should be noted that in reality the secondary star is not spherical but has the shape of on star's Roche lobe \citep[see e.g.][]{Shahbaz98}. In principle one should compute the orbital phase averaged spectrum taking into account the varying temperature and gravity across the secondary star’s Roche lobe (mainly due to the gravity darkening, inclination angle) by incorporating  synthetic spectra into the secondary star’s Roche geometry. However, it has been shown that phase averaging smooths out the effects of the phase-dependant Roche lobe rotation profile, and so the phase averaged spectrum can be modelled by the spectrum of a single temperature star convolved with the Gray rotation profile \citep{Gonzalez08}. Furthermore, it has been shown that LTE model atmospheres  adequately describe the atmospheres of highly irradiated companions in black widow MSPs, suggesting that the energy from the pulsar is being deposited deep enough within the companion such that the surface radiates as a black body in LTE \citep{Kennedy22}. Hence in the case of black widow MSP systems, 
these models are perfectly acceptable for modelling the secondary star.

We identify moderately strong, relatively isolated lines of several elements in the high-resolution solar flux atlas of \citep{Kurucz84}. We then inspect the spectrum of \target\ and select several spectral features containing Fe\,I and Ca\,I with excitation potentials between 1 and 5\,eV. We determine the synthetic spectrum for a given $T_{\rm eff}$, $\log\,g$, [Fe/H] combination and calculate the corresponding model atmosphere by interpolating the model atmosphere grid. Given that some of the Fe lines in our spectral range may be blended with Ca lines we also allow the [Ca/Fe] abundance to vary as a free parameter between $-$1.0 and $+$1.0\,dex. Note that the final determination of Ca abundance is done once the stellar parameters have been determined (see Section\,\ref{sec:abundance}). We also broaden the synthetic spectrum using the Gray rotation profile \citep{Gray92} with a linear limb darkening coefficient of 0.63, appropriate for a $\sim$G0V star \citep{Al-Naimiy78}. Finally, before computing the $\chi^2$ between the target and model spectrum, we degrade the synthetic spectrum using using a Gaussian instrumental profile and then bin the synthetic spectrum to the same velocity scale as the target spectrum.
To allow for the effects of a wavelength-dependent veiling caused by the presence of extra light in the system e.g. from an accretion disc \citep{Linares14,Shahbaz19} or intrabinary shock \citep{Romani16}, we subtract a scaled version of the synthetic spectrum from the observed spectrum. We model this wavelength-dependent scale factor as a linear function, where the $f_{\rm 6000}$ represents the fractional contribution of the secondary star to the observed spectrum at  6000\,\AA\ and  the gradient is given by $m_0$. The parameters $f_{\rm 6000}$ and $m_0$ range from 0.1 to 1.0 and $-$0.5$\times10^{-3}$ to $+$0.5$\times10^{-3}$ \AA$^{-1}$, respectively. 

\begin{figure*}
\subfloat[The disc-state spectrum.]{\includegraphics[width=0.50\linewidth,angle=0]{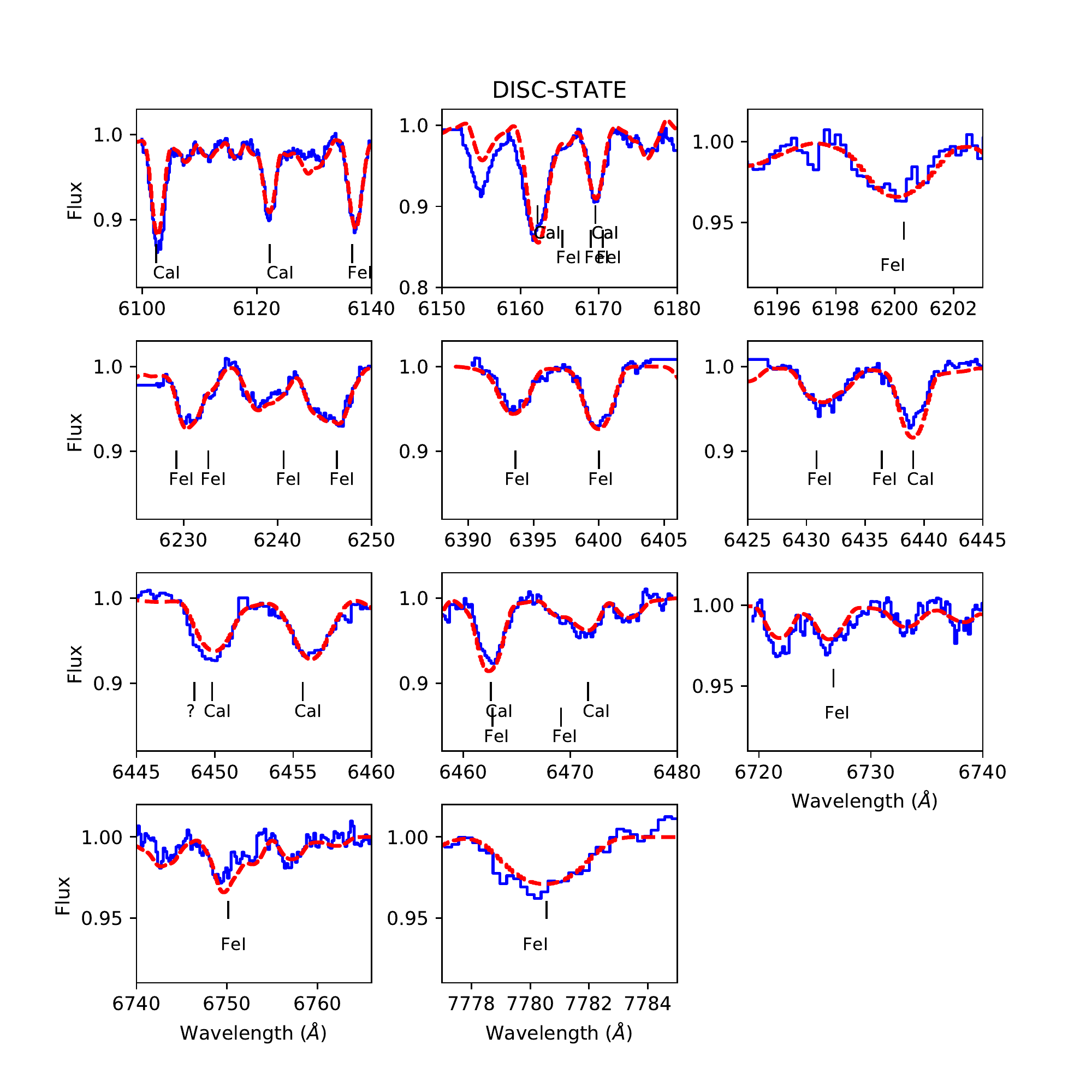}}
\subfloat[The pulsar-state spectrum.]{\includegraphics[width=0.50\linewidth,angle=0]{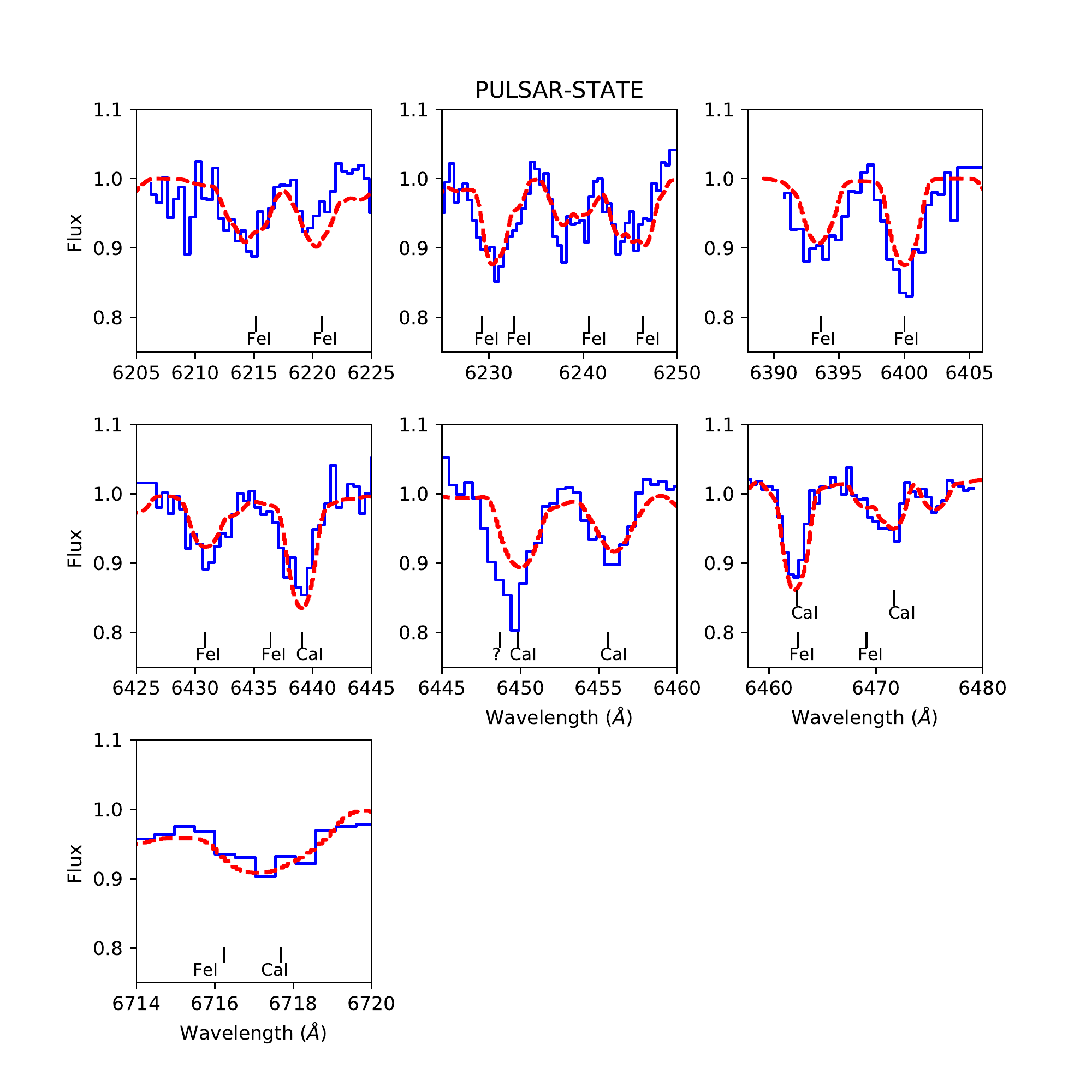}}
\caption{Sections showing the best-fit model (red dashed line) to the disc and pulsar-state spectrum (blue solid line) of the secondary star in  \target\, used to determine the stellar parameters. The stellar parameters are given in Table\,\ref{table:results} and Fig.\,\ref{fig:MCMC_stellar}
}
\label{fig:spec_DP}
\end{figure*}

We use five small spectral regions with a width of 20\,\AA\ centred on 6228\,\AA, 6435\,\AA\ and 6726\,\AA\ for the pulsar-state spectra and 6140\,\AA, 6222\,\AA, 6435\,\AA, 6745\,\AA\ and 7749\,\AA\ for the disc-state spectra regions  devoid of emission lines (see Fig.\,\ref{fig:spec_DP}). The model free parameters for each state are therefore, $T_{\rm eff}$, $\log\,g$ $f_{\rm 6000}$, $m_{\rm 0}$, [Fe/H], [Ca/Fe] and a wavelength shift (to allow for any possible shift between the observed and model spectrum).

The surface gravity of the secondary star is determined by its mass ($M_{\rm 2}$) and radius ($R_{\rm 2}$) as $\log g = 4.438 + \log M_{\rm 2} -2\log R_{\rm 2}$.We can express $R_{\rm 2}$  in terms of the star's equivalent volume radius \citep{Eggleton83} $r_{\rm eqv}$ (in units of the binary separation, $a$), which depends on the binary mass and Roche lobe filling factor, $f$ \citep{Shahbaz17}. Given that $M_{\rm 2}$ = q$M_{\rm 1}$ and $a = 4.207 M_{\rm 1}^{1/3} (1+q)^{1/3} P_{\rm orb}^{2/3}$\Rsun, the star's gravity can be expressed as 

\begin{equation}
\log g = 4.438 + q\log M_{\rm 1} - 2\log a - 2\log r_{\rm eqv}(f,q) 
\end{equation}

\noindent 
Using $q$ = 0.137 \citep{Shahbaz19} and the expected limits on $M_{\rm 1}$ = 1.4--3.0\Msun (from a canonical to the maximum neutron star mass), we find that in the disc-state ($f$ = 1), $\log\,g^{\rm DS}$ is expected to lie in the range 4.57 to 4.68. Note that in the pulsar-state the secondary star may slightly underfill its Roche lobe\citep{Stringer21}, but should be close to unity in the disc-state (because an accretion disc is observed). This means that the radius of the secondary star is smaller in the pulsar-state compared to the disc-state. We can therefore impose the following conditions: (1) the secondary star's rotational broadening in the disc-state is larger than the star's rotational broadening in the pulsar-state, (2) $\log\,g$ in the pulsar-state is larger than $\log\,g$ in the disc-state and (3) $\log\,g$ in the disc-state lies in the range 4.57 to 4.68.

We perform the spectral fitting with the Markov-chain Monte Carlo (MCMC) package \textsc{emcee} \citep{Foreman13}, using the $\chi^2$ minimization method and assuming flat priors. A total of 50 walkers were used with 5000 samples taken and a burn-in of 100 samples. From our preliminary fits we find that $f_{\rm 6000}$=1 is required for the pulsar-state spectra. For speed, we therefore repeat the MCMC fits but fixing $f_{\rm 6000}$=1 for the pulsar-state spectrum. For both the disc- and pulsar-state spectrum we find that we cannot constrain the  gravity of the secondary star. In Fig.\,\ref{fig:MCMC_stellar} we show the resulting relevant MCMC 2-D model parameter distributions. Inspection of the posteriors of the parameters do not show any obvious correlations between the parameters, and the individual parameter distributions follow normal distributions. The most likely values for the constrained parameters are given in Table\,\ref{table:results}. The sections of the disc- and pulsar-state spectrum along with the best model fit is shown in Fig.\,\ref{fig:spec_DP}.

As a test we also perform the same analysis on a template field star taken from our UVES Paranal Observatory Project \citep{Bagnulo03} library of high signal-to-noise template spectra \citep{Shahbaz19}.  We rotationally broaden the template star spectrum HD\,115617 by 78 \kms. We then degrade the template star spectrum to match the same spectral resolution and heliocentric velocity scale as the target spectrum in the pulsar- and disc-state. Finally we determine the stellar parameters using the same method and spectral regions as the target spectrum. We obtain the stellar parameters 
$T_{\rm eff}$ = 5470\,K, $\log\,g$ = 4.5 $\rm cm\,s^{-2}$ and [Fe/H] = 0.06\,dex
which are consistent with the observed value in the SIMBAD database \citep{Wenger00}.

\subsection{Stellar Abundances}
\label{sec:abundance}

\begin{figure*}
\subfloat[The Ca lines in the 6425--6470\,\AA\ disc-state spectrum.]{\includegraphics[width=0.40\linewidth]{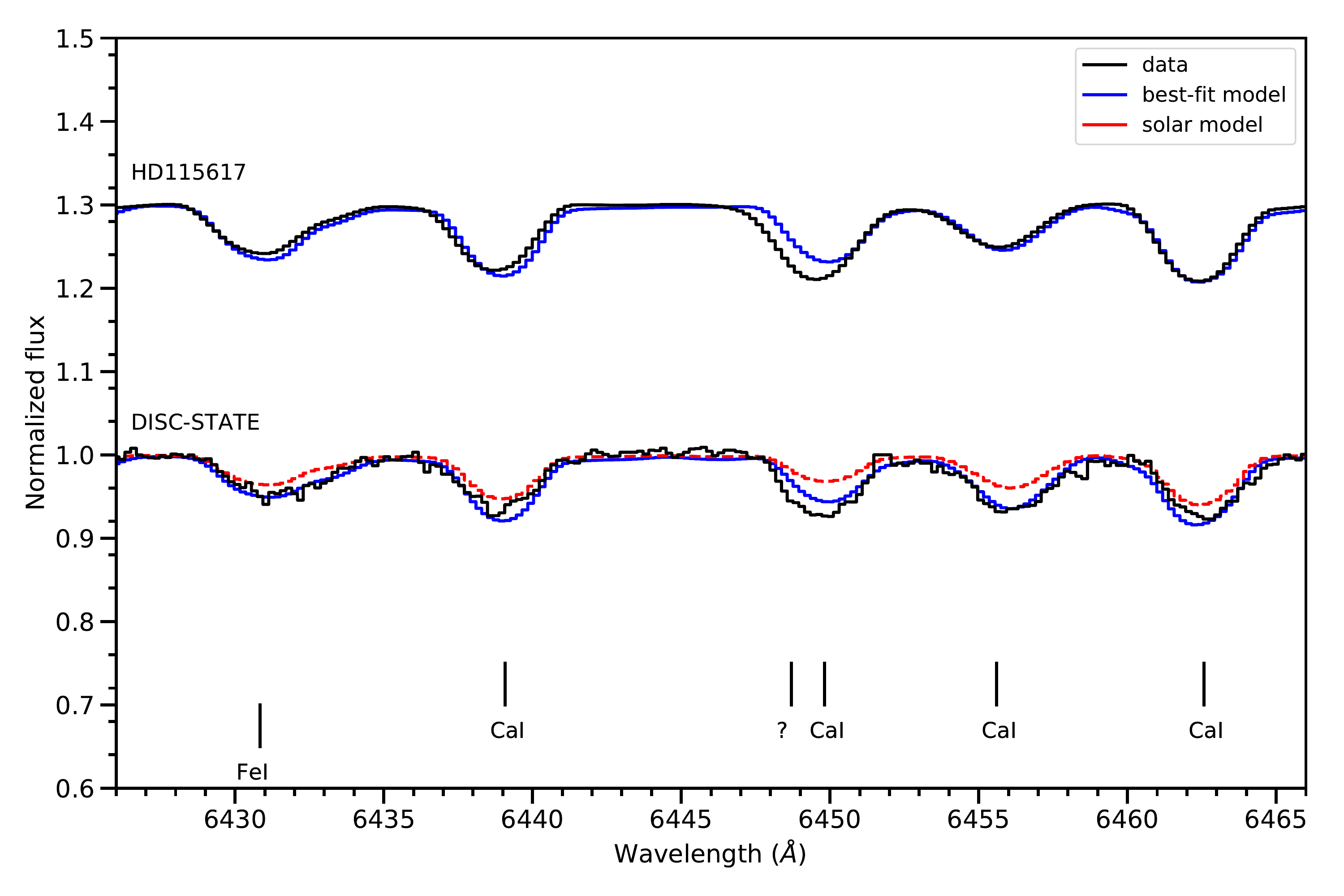}\label{fig:spectrum_a}}
\subfloat[The Ca lines in the 6235--6265\,\AA\ pulsar-state spectrum.]{\includegraphics[width=0.40\linewidth]{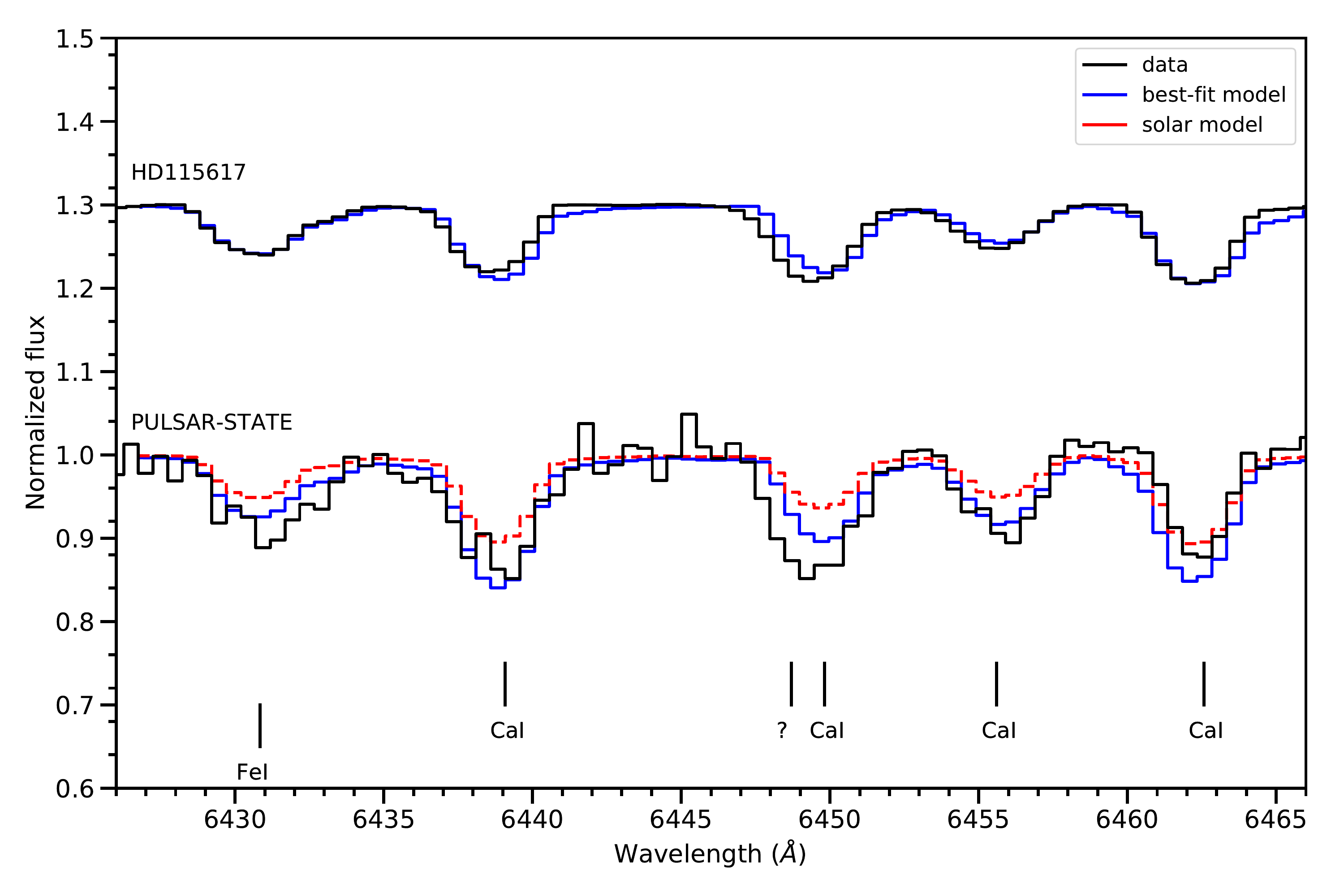}\label{fig:spectrum_b}}

\subfloat[The Si lines in the 6235--6265\,\AA\ disc-state spectrum.]{\includegraphics[width=0.40\linewidth]{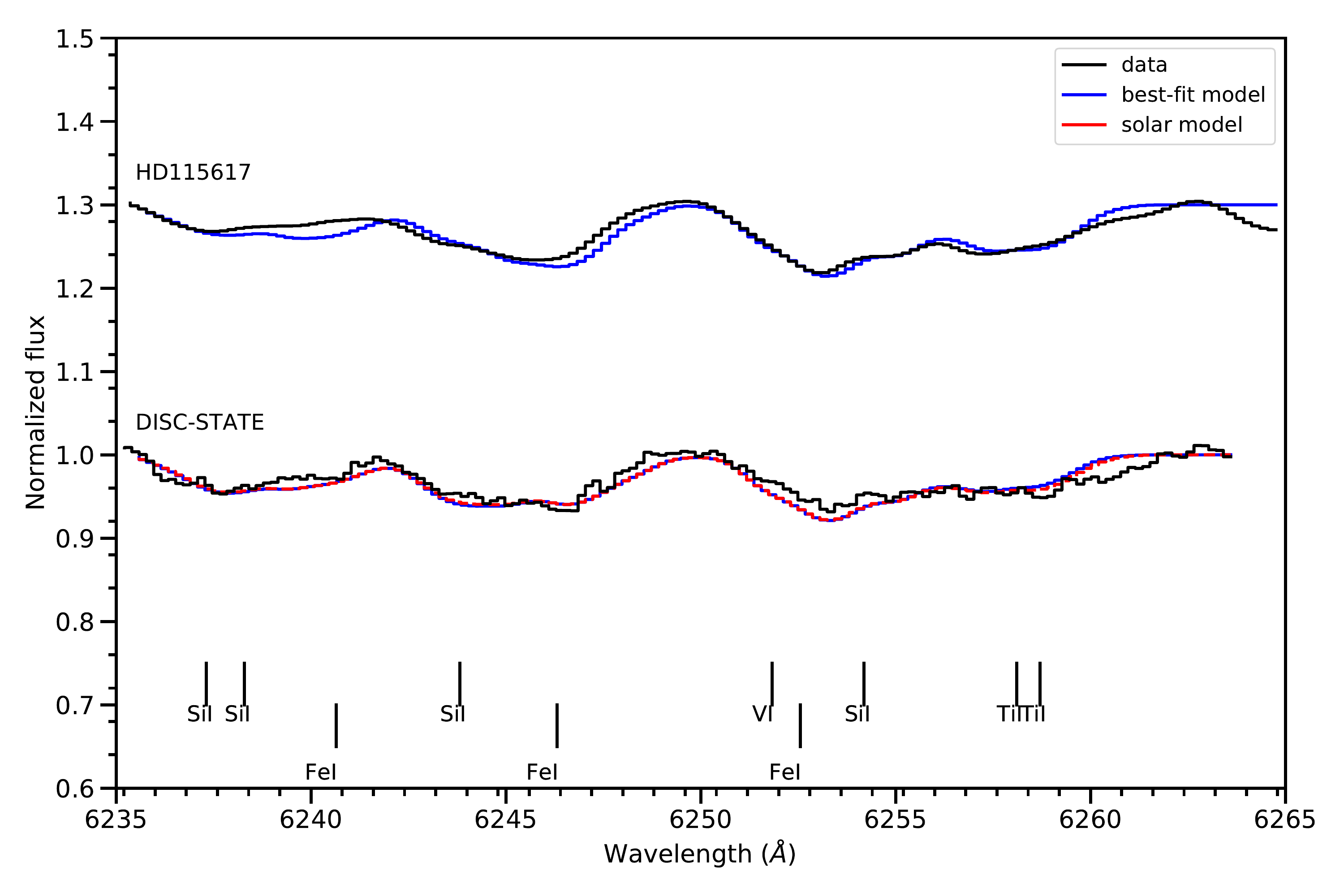}\label{fig:spectrum_c}}
\subfloat[The Si lines in the 6235--6265\,\AA\ pulsar-state spectrum.]{\includegraphics[width=0.40\linewidth]{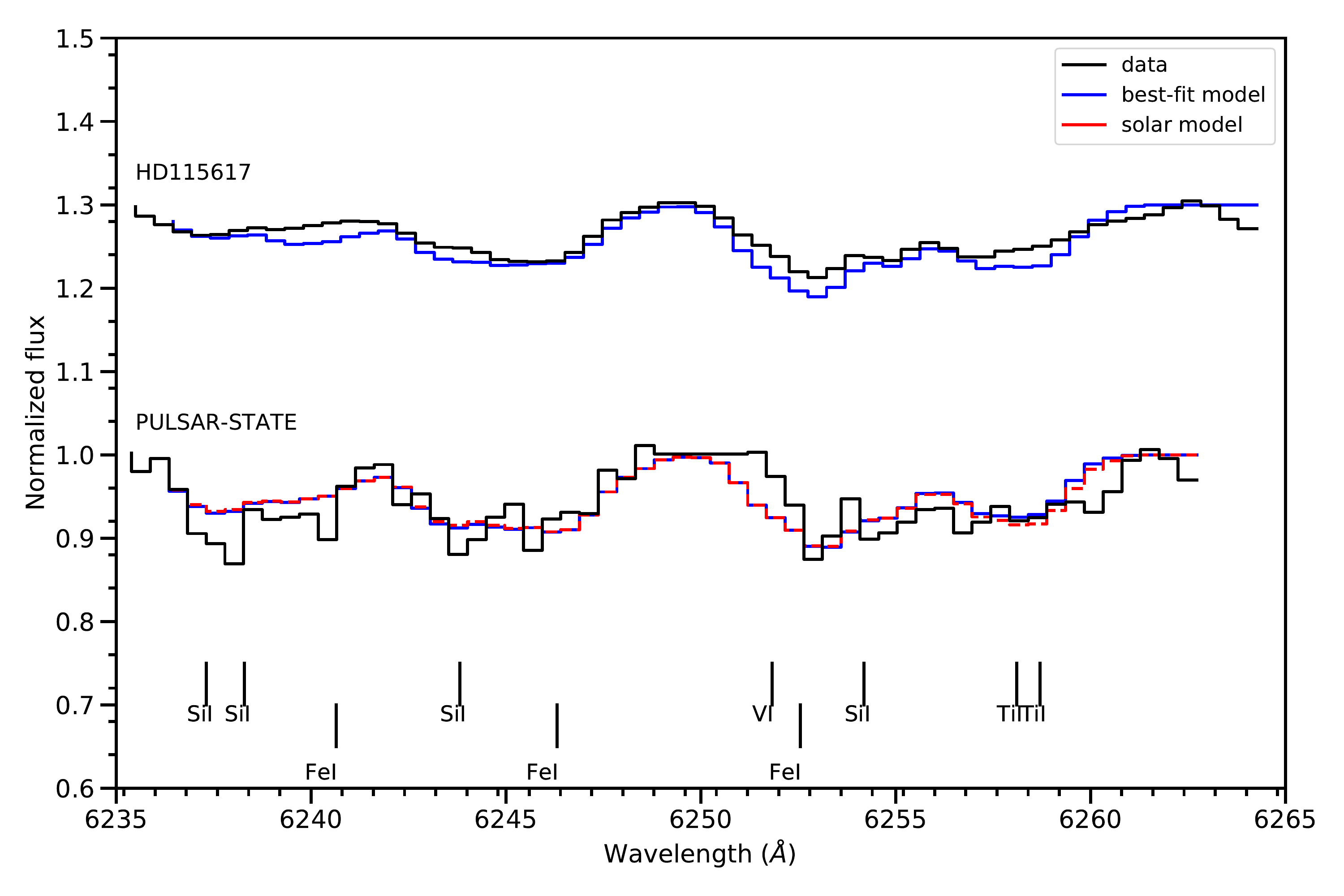}\label{fig:spectrum_d}}

\subfloat[The Al and Li lines in the  6690--6730\,\AA\ pulsar-state spectrum.]{\includegraphics[width=0.40\linewidth]{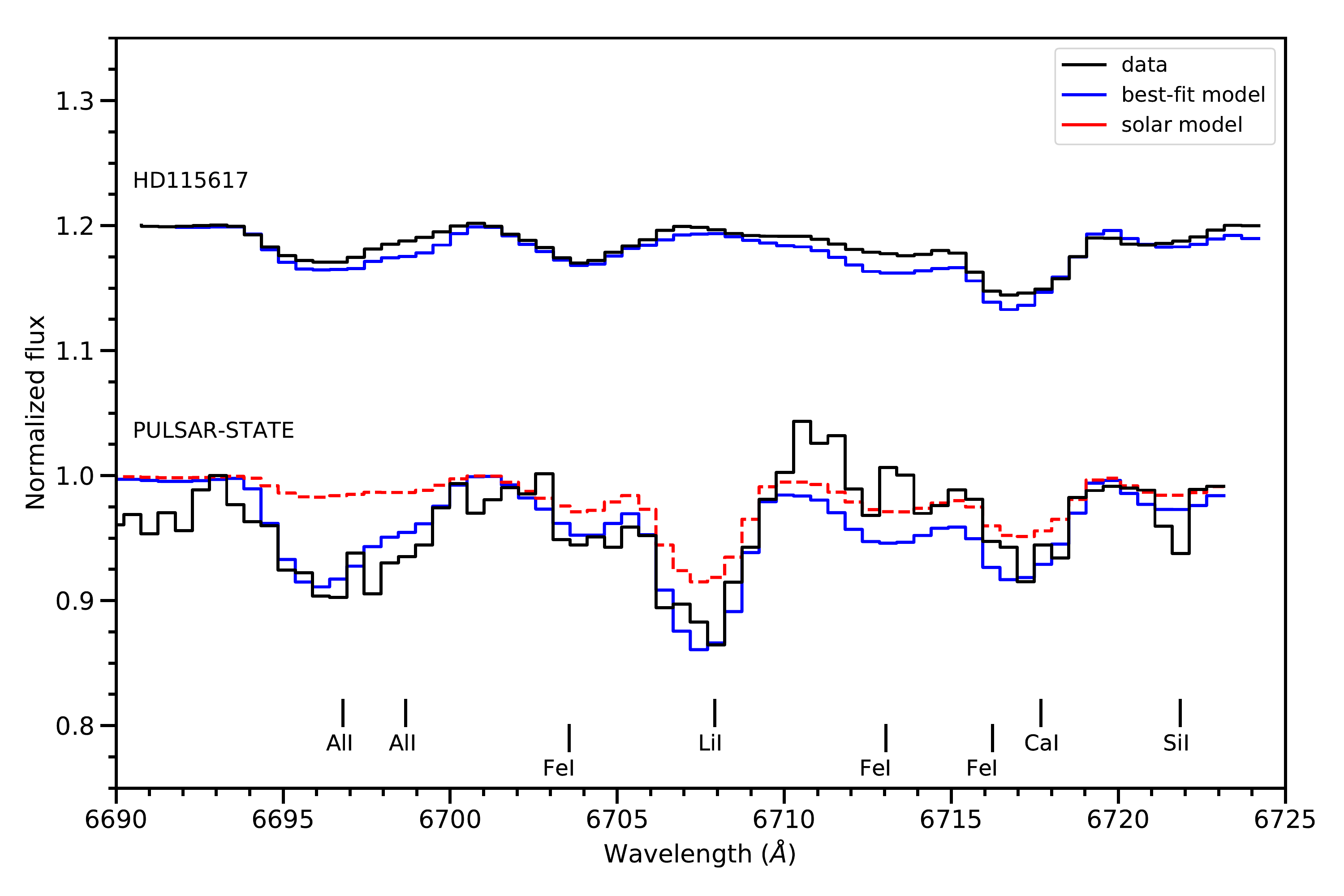}\label{fig:spectrum_e}}
\subfloat[The O and Ni lines in the 7765--7795\,\AA\ disc-state spectrum.]{\includegraphics[width=0.40\linewidth]{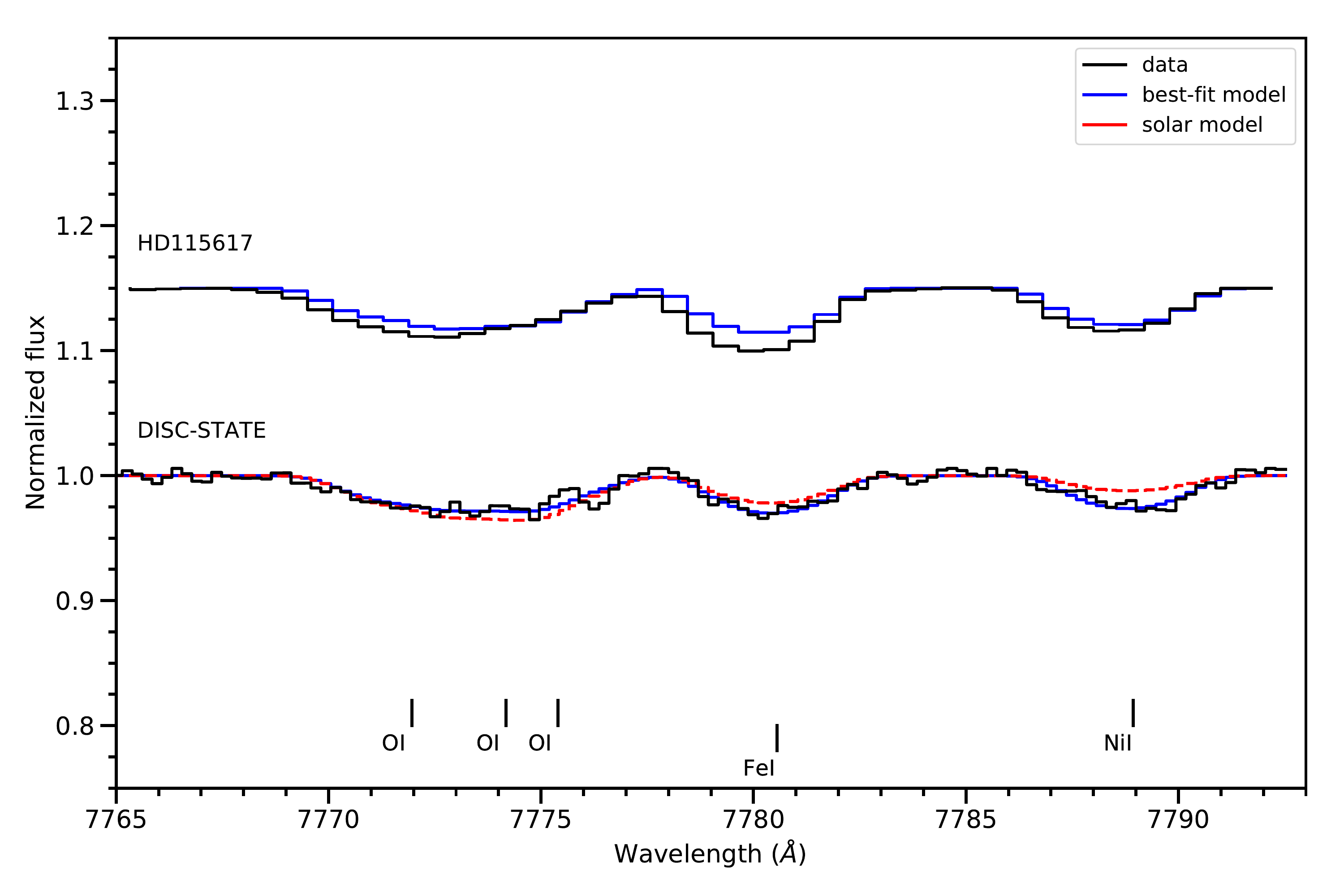}\label{fig:spectrum_f}}
\caption{The best synthetic model fit synthetic model to the pulsar and disc-state spectrum of the secondary star in \target. In each plot we also show fits to a G6 template star. The synthetic spectrum for Solarabundances (red dotted line) and best-fit abundances (blue solid line) are shown. }
\label{fig:spectrum}
\end{figure*}

\begin{figure}
\centering
\includegraphics[width=1.0\linewidth,angle=0]{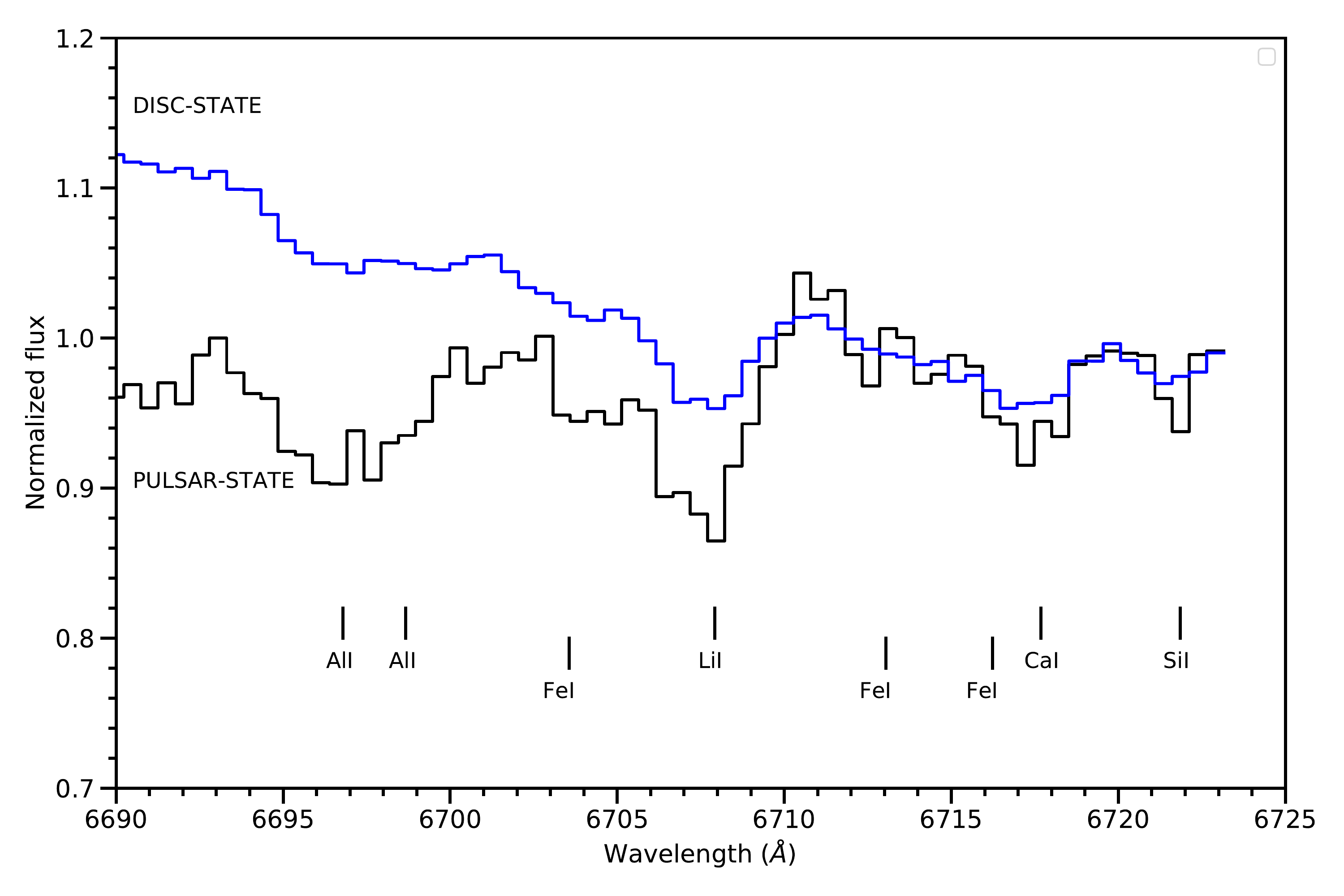}
\caption{
The pulsar- and disc-state (blue) spectra in the 6690--6730\,\AA\ spectral region showing the Al\,I 6696\,\AA\ and Li\,I 6708\,\AA\ doublets. Note the contamination of the disc-state spectrum from the broad He\textsc{II} 6678 \AA\ emission line arising from the accretion disc.}
\label{fig:Li_P_D}
\end{figure}

Using the derived stellar parameters we identify several spectral regions containing various lines of O, Al, Si, Ca, Ni and Li, some of which are blended with Fe lines. We determine the abundances of these elements by comparing the observed pulsar- and disc-state spectra with modified versions of synthetic spectra computed using the best model atmosphere calculated with the corresponding stellar parameters and metallicity determined in Section\,\ref{sec:stellar}. Again, before comparison we rotationally broaden and degrade the model spectrum to the same velocity scale as the target spectrum.  We perform an MCMC spectral analysis 
\citep{Foreman13} using the $\chi^2$ minimization method and assume Gaussian priors  for the corresponding stellar parameters and metallicity, and flat priors for the element abundances. The abundance [X/H] was allowed to vary between -2.0 and 2.0\,dex.
The Ca (see Figs.\,\ref{fig:spectrum_a} and \ref{fig:spectrum_b}) and Si (see Figs.\,\ref{fig:spectrum_c} and \ref{fig:spectrum_d})  abundances  are derived from  several lines in the spectral region 6425--6470\,\AA\ and 6235--6260\,\AA,  respectively in both the disc- and pulsar-state spectra.
Given the  presence of H and He emission lines from the accretion disc in the disc-state, the Al and Li abundances are determined only from the pulsar-state spectrum by simultaneously fitting the Al\,I 6696\,\AA\ and Li\,I 6708\,\AA\ doublets (see Fig.\,\ref{fig:spectrum_e}). In Fig.\,\ref{fig:Li_P_D} we show both the pulsar- and disc-state spectra, where the contamination from the broad He\,II 6678.149\AA\ emission can be clearly seen in the disc-state spectrum. 

\begin{figure}
\centering
\begin{tabular}{c}
\includegraphics[width=0.8\linewidth,angle=0]{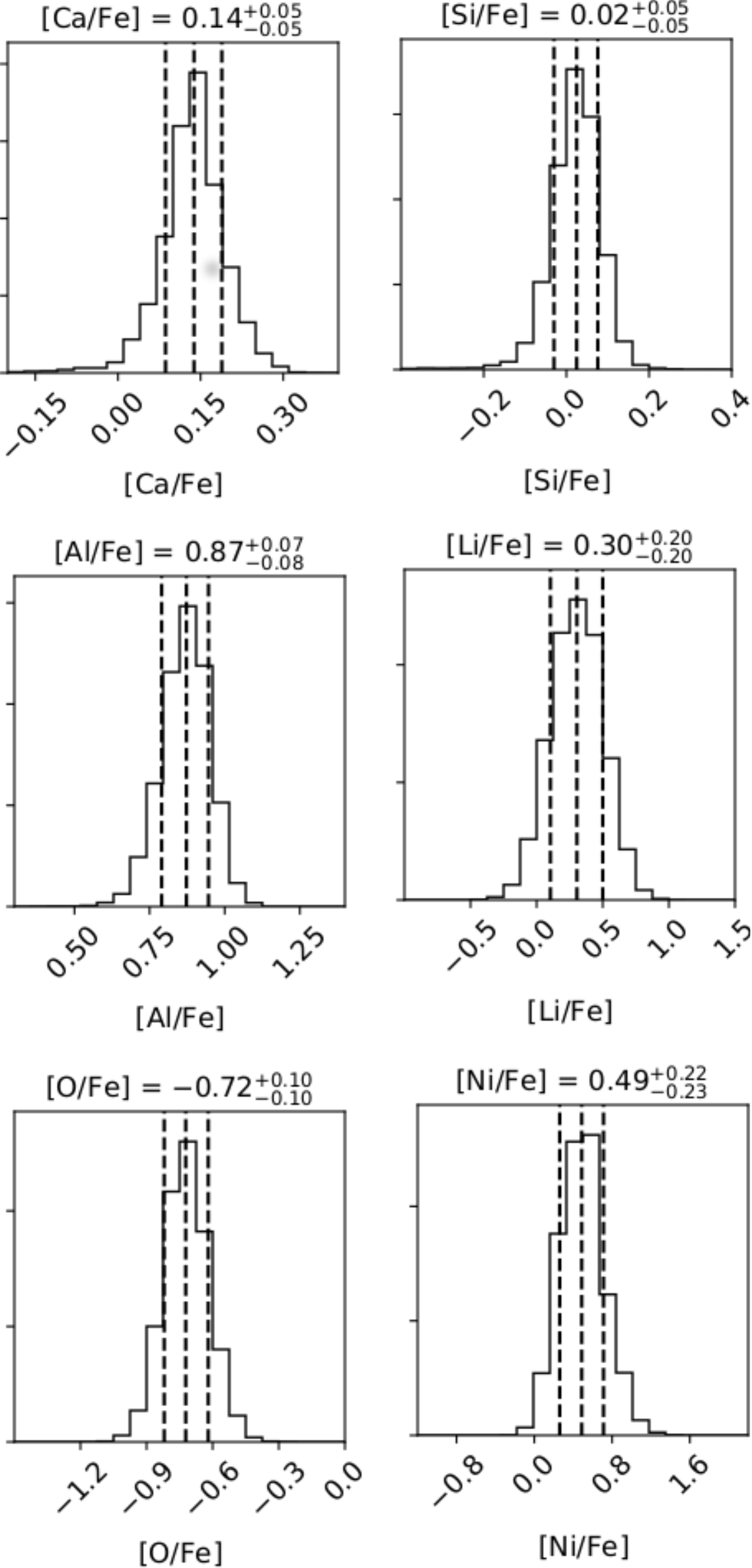}
\end{tabular}
\caption{The results of the MCMC model parameter distributions resulting from synthetic fits to the observed spectrum of \target\ to determine the abundance of various elements. The projected 1-D parameter distributions with the mean  and standard deviation are given in each plot.}
\label{fig:MCMC_X}
\end{figure}

\begin{figure}
\centering
\includegraphics[width=1.0\linewidth,angle=0]{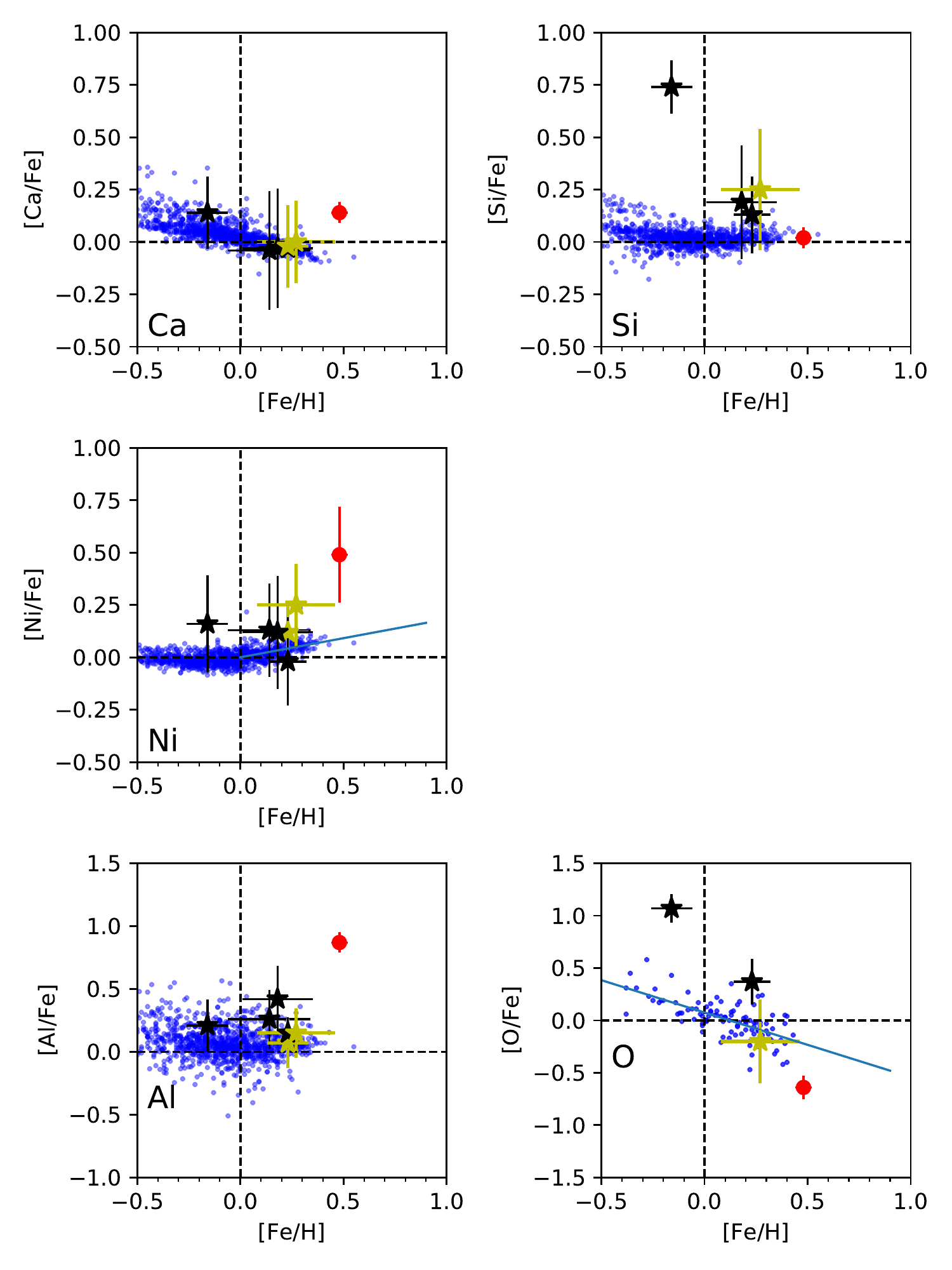}
\caption{
Abundance ratios of the secondary star in black hole  (black stars) and  neutron star (yellow stars) X-ray binaries (see Table\,\ref{table:Li_XRBs}) in comparison with the abundances of solar-type stars in the Solar neighbourhood (blue circles). The red circles show the abundance ratio in \target. Galactic trends are taken from \citet{Adibekyan12} and \citet{Ecuvillon06}. [O/Fe] has been corrected for NLTE effects.
} 
\label{fig:Fe-X}
\end{figure}

The O and Ni abundances are derived using the O\,I 7771\,\AA\ triplet and  Ni\,I 7790\,\AA\ lines, respectively, and  are fitted simultaneously using the disc-state spectrum only, as the pulsar-state spectrum do not cover this spectral range (see Fig.\,\ref{fig:spectrum_f}).
The results of the MCMC analysis are shown in Fig.\,\ref{fig:MCMC_X}.
We also perform the same analysis on the template star spectrum and the corresponding plots are shown in Fig.\,\ref{fig:spectrum}. The abundances determined for the various  elements are given in Table\,\ref{table:X}. In Fig.\,\ref{fig:Fe-X} we show the abundances of the secondary star in \target\ in comparison with the abundances in G and K metal-rich dwarf stars \citep{Adibekyan12,Ecuvillon06}.

\begin{table*}
\centering
\caption{LTE element abundances of black hole and neutron star X-ray binaries}
\begin{tabular}{lrccrrccc}
\hline
Target          &  [O/H]$^{\rm a}$     &   [Al/H]        &  [Si/H]         &  [Ca/H]         &  [Fe/H]         &  [Ni/H]         &  A(Li)$^{\rm a,b}$	 & Ref.$^{\rm c}$ \\
\hline
\multicolumn{9}{c}{Black hole systems}  \\
A0620--00       &		         &  0.40 $\pm$ 0.12  &               &  0.10 $\pm$ 0.20  &  0.14 $\pm$ 0.20  &  0.27 $\pm$ 0.10  &  2.31 $\pm$ 0.21  &  1   \\
XTE\,J1118+480  &		         &  0.60 $\pm$ 0.20  &               &  0.15 $\pm$ 0.23  &  0.18 $\pm$ 0.17  &  0.30 $\pm$ 0.21  &  $<$1.8	         &  4   \\
Nova\,Sco\,94   &  0.91 $\pm$ 0.09 &  0.05 $\pm$ 0.18  &  0.58$\pm$0.08  & $-$0.02 $\pm$ 0.14  & $-$0.11 $\pm$ 0.09  &  0.00 $\pm$ 0.21  &  $<$2.16	         &  5   \\
V404\,Cyg       &  0.60 $\pm$ 0.19 &  0.38 $\pm$ 0.09  &  0.36$\pm$0.11  &  0.20 $\pm$ 0.16  &  0.23 $\pm$ 0.09  &  0.21 $\pm$ 0.19  &  2.70 $\pm$ 0.40  &  6   \\ \\
\multicolumn{9}{c}{Neutron star systems}  \\
Cen\,X--4       &		         &  0.30 $\pm$ 0.17  &               &  0.21 $\pm$ 0.17  &  0.23 $\pm$ 0.10  &  0.35 $\pm$ 0.17  &  2.98 $\pm$ 0.29  &  2   \\
Cyg\,X--2       &    0.07 $\pm$ 0.35 &  0.42 $\pm$ 0.05  &  0.52 $\pm$ 0.22  &  0.27 $\pm$ 0.05  &  0.27 $\pm$ 0.19  &  0.52 $\pm$ 0.05  &  $<$1.48	         &  3   \\
PSR\,J1023+0038 & $-$0.64 $\pm$ 0.11 &  1.35 $\pm$ 0.09  &  0.50 $\pm$ 0.06  &  0.62 $\pm$ 0.06  &  0.48 $\pm$ 0.04  &  0.97 $\pm$ 0.23  &  3.66 $\pm$ 0.20  &  7    \\
\hline
\end{tabular}
\begin{tablenotes}
  \item  $^{\rm a}$ Corrected for NLTE effects, 
  \item  $^{\rm b}$ Expressed as $\rm A(Li)$  = $\log$ [$N$(Li)/$N$(H)]] + 12.
  \item  $^{\rm c}$ References: $(1)$ \citet{A0620}; $(2)$ \citet{CenX4}; $(3)$ \citet{CygX2}; $(4)$ \citet{J1118_2}; $(5)$ \citet{J1655}; $(6)$ \citet{V404Cyg,Martin92}; $(7)$ This work  
\end{tablenotes}
\label{table:Li_XRBs}
\end{table*}

\section{Results}

\subsection{Stellar parameters}

The light curves taken when \target\ was in the pulsar- \citep{Woudt04,TA05} and disc-state \citep{Bogdanov14,Shahbaz15} show an asymmetric single-humped modulation. In the pulsar-state this is due to the combined effects of the tidally-locked secondary star's ellipsoidal modulation and the high-energy emission from the pulsar's relativistic wind powered by the rotational spin down of the neutron star heating the inner face of the secondary star. In the disc-state, as well as the star's ellipsoidal modulation, the observed X-ray and gamma-ray luminosity (most likely due to accretion)  is sufficient to provide the observed irradiating luminosity. Irradiation also has a pronounced effect on the phase-resolved spectra, where spectral type changes along the orbit have been observed \citep{Shahbaz19}. The pulsar- to disc-state transition involves an increase in the gamma-ray and X-ray flux of a factor of $\sim$20 \citep{Patruno14,Takata14}, which implies that during the disc-state the  secondary star is subject to irradiation from the accretion disc.

We determine the global temperature of the secondary star in the disc- and pulsar-state to be $T_{\rm eff}^{\rm PS}$ = 5724\,K and $T_{\rm eff}^{\rm DS}$ = 6128\,K, respectively, which correspond to spectral types of $\sim$G1 and F8, respectively \citep{Pecaut13}. The transition from the pulsar- to disc-state results in a significant (3.8-$\sigma$ level) increase in temperature, caused by the increased gamma-ray and X-ray flux. In contrast, from Kepler photometry \citealt{Kennedy18} find that the heating amplitude does change between states. However, it should be noted that absorption line spectroscopy is more  sensitive to changes in \teff\ than broad-band photometry.

\subsection{Chemical abundances}

The Ca and Ni abundances were determined from features with little blending from other elements, mostly Fe lines. Several Ca spectral features are shown in Fig.\,\ref{fig:spectrum}a. In general, these features are well reproduced by the synthetic spectra, except for one feature ($\sim$6450\,\AA), which was blended with an unidentified line in the Solar spectrum and which was not used in the chemical analysis. In Fig.\,\ref{fig:spectrum}b we show Si lines used in the abundance analysis. In Fig.\,\ref{fig:spectrum}c  we show the region which contain the Al and Li doublets, and in Fig.\,\ref{fig:spectrum}d we show the O and Ni features.

The [Fe/H] ratio of 0.48 $\pm$ 0.04 determined is surprisingly high, $\sim$2 times higher than that found in neutron star X-ray binaries \citep{V404Cyg}. In Fig.\,\ref{fig:Fe-X} we show the element abundances of the secondary star in  \target\  relative to Fe in comparison with the Galactic abundance trends \citep{Adibekyan12}  as well as the element abundances of the secondary stars in X-ray binaries \citep{V404Cyg}. We find that in general the abundances of the secondary star in \target\ are different to the elemental abundance of the secondary stars in other X-ray binaries and to stars in the Solar neighbourhood. The abundances ratios of Ca and Al, with respect to Fe in the secondary star of \target\ are  higher than those in stars of similar Fe content (see Fig.\,\ref{fig:Fe-X}) whereas, the Si abundance  seems to be consistent with the Galactic trends of stars in the Solar neighbourhood. Ni and Al with respect to Fe are  clearly enhanced compared to stars in the Solar neighbourhood. 

\subsubsection{Oxygen}

The O abundance was derived from  the O\,I near-infrared triplet O\,I 7771\,\AA\ in the disc-state spectrum (see Fig.\,\ref{fig:spectrum}) which is well reproduced by the synthetic spectra. The O\,I feature in the secondary star is under-enhanced compared with a template star having similar effective temperature (Solar abundance model). We obtain a best-fit LTE abundance ratio of [O/Fe]$_{\rm LTE}$ = $-$0.74 $\pm$ 0.11\,dex, which corresponds to [O/H]$_{\rm LTE}$ = $-$0.26$\pm$0.12\,dex. The O\,I 7771\,\AA\  triplet suffers from significant non-LTE (NLTE) effects \citep{Ecuvillon06}. For the stellar parameters and O abundance of the secondary star in \target, the NLTE correction\footnote{$\rm \Delta NLTE = A(X)_{\rm NLTE} - A(X)_{\rm LTE}$} is estimated to be $\sim -$0.40\,dex. Applying this correction to the LTE O abundance leads to [O/H]$_{\rm NLTE} \sim-$1.14\,dex, which implies that O is significantly depleted. In Table\,\ref{table:X}  we give the O abundance corrected for NLTE effects. The secondary star in \target\ has an anomalously low abundance ratio [O/Fe]$_{\rm NLTE} \sim -$1.08\,dex which is clearly not consistent with the Galactic trend  measured in stars in the Solar neighbourhood (see Fig.\,\ref{fig:Fe-X}).

\subsubsection{Lithium}

\begin{figure}
\centering
\includegraphics[width=1.0\linewidth,angle=0]{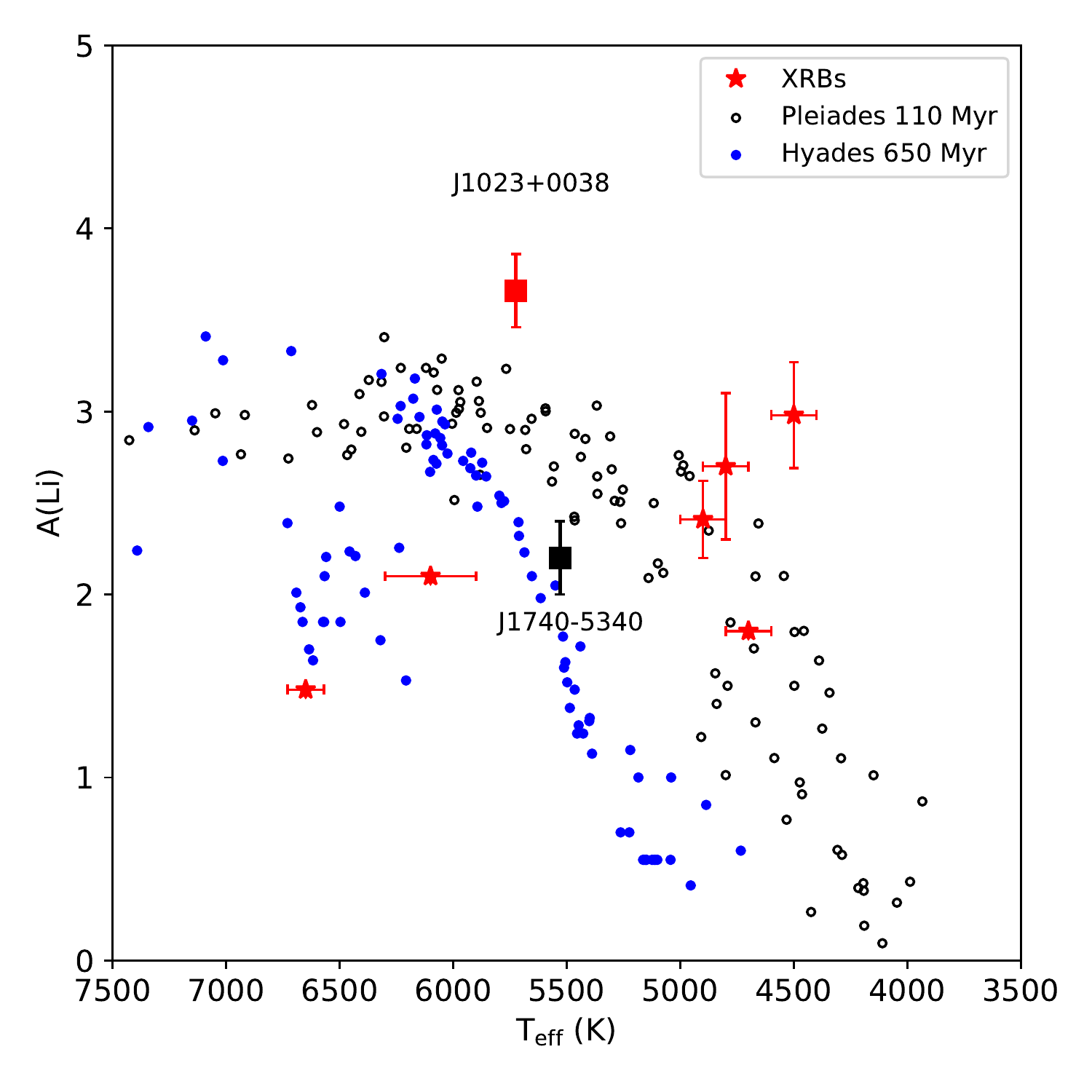}
\caption{
Li abundance of the secondary star in \target\ (filled blue circle) in comparison with the secondary stars in X-ray binaries (red filled stars) and stars in the young clusters, Pleiades \citep{Barrado16} and Hyades \citep{Cummings17}.
} 
\label{fig:Li}
\end{figure}

The best-fit to the Li\,I 6708\,\AA\ feature gives an LTE abundance of $\rm A(Li)_{\rm LTE}$ = 4.09 $\pm$ 0.20\,dex. 
Using 3D non-LTE model atmospheres corrections \citep{Wang21}, 
calculated using \textsc{Breidablik} \footnote{https://github.com/ellawang44/Breidablik} 
we find  $\rm \Delta NLTE \sim$ -0.43\,dex, which implies
$\rm A(Li)_{\rm NLTE}$ = 3.66 $\pm$ 0.20\,dex for the secondary star in \target. 
For the template star spectrum HD\,115617 we observe no Li and obtain results consistent with \citep{Gonzalez10b}.
In Fig.\,\ref{fig:Li} we compare the Li abundance of the secondary star in \target\ with the secondary stars in X-ray binaries and with stars in the Pleiades and Hyades cluster. The secondary star in \target\ has a Li abundance higher than  late-type main sequence stars as well as X-ray binaries (see Section\,\ref{sec:comparison}). 

\section{Comparison with young clusters and X-ray binaries}
\label{sec:comparison}

The secondary star in \target\ shows an unexpectedly high Li abundance, higher than the cosmic value seen in stars in the Galactic disk. Convective mixing is expected to deplete Li in the atmosphere of a G0 star during its pre-main sequence and main-sequence evolution. The mean Li abundance of stars in the 5700--6100\,K temperature range decrease with age. This effect can be clearly seen in young open clusters such as the Pleiades cluster \citep[$\sim110$\,Myr;][]{Dahm15} and the Hyades cluster \citep[$\sim650$\,Myr;][]{Martin18}. In Fig.\,\ref{fig:Li}, we compare the Li abundance in \target\ with stars in the Pleiades and Hyades clusters. Clearly the Li content in the secondary star is too high compared to stars with the same effective temperature, which indicates that in absence of a Li production mechanism, the secondary star in \target\ is much younger than the Pleiades cluster. 
\citet{Franciosini20} have studied the Li abundances of some young open clusters showing that the mean Li abundance of G0 stars decreases from A(Li) $\sim$ 3.6--3.8\,dex in $\gamma$Velorum (10--20\,Myr) to 3.2\,dex in NGC2547 (35--45\,Myr) down to about 3.0\,dex in NGC\,2516 (70--150\,Myr) as in the Pleiades cluster \citep{Barrado16}. \citet{Lim16} measured Li abundance in a significant number of stars in the cluster NGC\,2264 with \teff = 4000--6500\,K, confirming that the initial Li abundance at ages 3--5\,Myr is already around the cosmic value (A(Li) $\sim$ 3.2 $\pm$ 0.2\,dex).  This demonstrates that the Li abundance of the secondary star in \target\ is anomalously high, irrespective of the age of the system. There a few cases of unevolved stars (in this context, stars which have not reached the red giant branch; RGB) with extremely high Li content. One interesting example is a turn-off star \#1657 in the metal-poor globular cluster NGC\,6397, which shows A(Li) $\sim$ 4.1\,dex \citep{Koch11}, i.e. about 100 times higher Li content 
compared to the primordial Li abundance produced in the Big Bang. This result remains puzzling and unusual scenarios are involved in order to explain the strong Li enhancement, such as the capture of a substellar body, type-II supernovae, diffusion, contamination from an AGB/RGB companion, some of them excluded by the normal abundance of Be \citep{Pasquini14}.

In general the abundances of heavy elements in the secondary star of \target\ are higher than the Solar value, and indeed higher than those of the most metal-rich stars of the Solar neighbourhood. In Fig.\,\ref{fig:Fe-X} we show the element abundances of the secondary star in \target\ relative to Fe in comparison with the Galactic abundance trends \citep{Adibekyan12} as well as the element abundances of secondary stars in other X-ray binaries \citep{V404Cyg}. The abundances ratios of [Ca/Fe], [Al/Fe] and [Ni/Fe] are higher than those in stars of similar Fe content (see Fig.\,\ref{fig:Fe-X}), [O/Fe] is surprisingly under-abundant and [Si/Fe] seems to be consistent with the Galactic trends of stars in the Solar neighbourhood. 

The metallicity in the secondary star of \target\ ([Fe/H]$\sim$0.48) is much higher than the metallicity of other secondary stars in the neutron star X-ray binaries Cen\,X-4 and Cyg\,X-2 \citep{CenX4,Suarez-Andres15}. In Cen\,X-4 and Cyg\,X-2, Ni is also over-abundant and in Cyg\,X-2 O appears to be under-abundant \citep{Suarez-Andres15}. Indeed, the average metallicity of secondary stars in neutron star X-ray binaries ([Fe/H]$\sim$0.24) seems to be higher than in black hole X-ray binaries ([Fe/H]$\sim$0.09) \citep{V404Cyg}, suggesting that a high amount of Fe is always ejected during SN explosions that forms the NS. 
The Li abundance is higher than field late-type main sequence stars as well as X-ray binaries (see Table\,\ref{table:Li_XRBs}). 

\section{Chemical anomalies due to supernova}
\label{sec:SN}

It is generally believed that during the common envelope evolution of an interacting binary, a compact object (neutron star or black hole) is subsequently born after the explosion of the He star. 
Part of the ejected mass in the SN explosion may be captured by the secondary star, leading to abundance anomalies as for instance in the case of the black hole X-ray binary Nova\,Sco\,1994 \citep{Israelian99,J1655}. The chemical abundances of the secondary stars in X-ray binaries show over-abundance of several $\alpha$-elements (such as O, S, Si, Ca). This has been interpreted as pollution by matter ejected during the SN/HN \citep[see][and references within]{V404Cyg}. In the following we consider this scenario for \target.


\begin{figure}
\subfloat[Spherically symmetric core-collapsed SN explosion.]{\includegraphics[width=0.95\linewidth,angle=0]{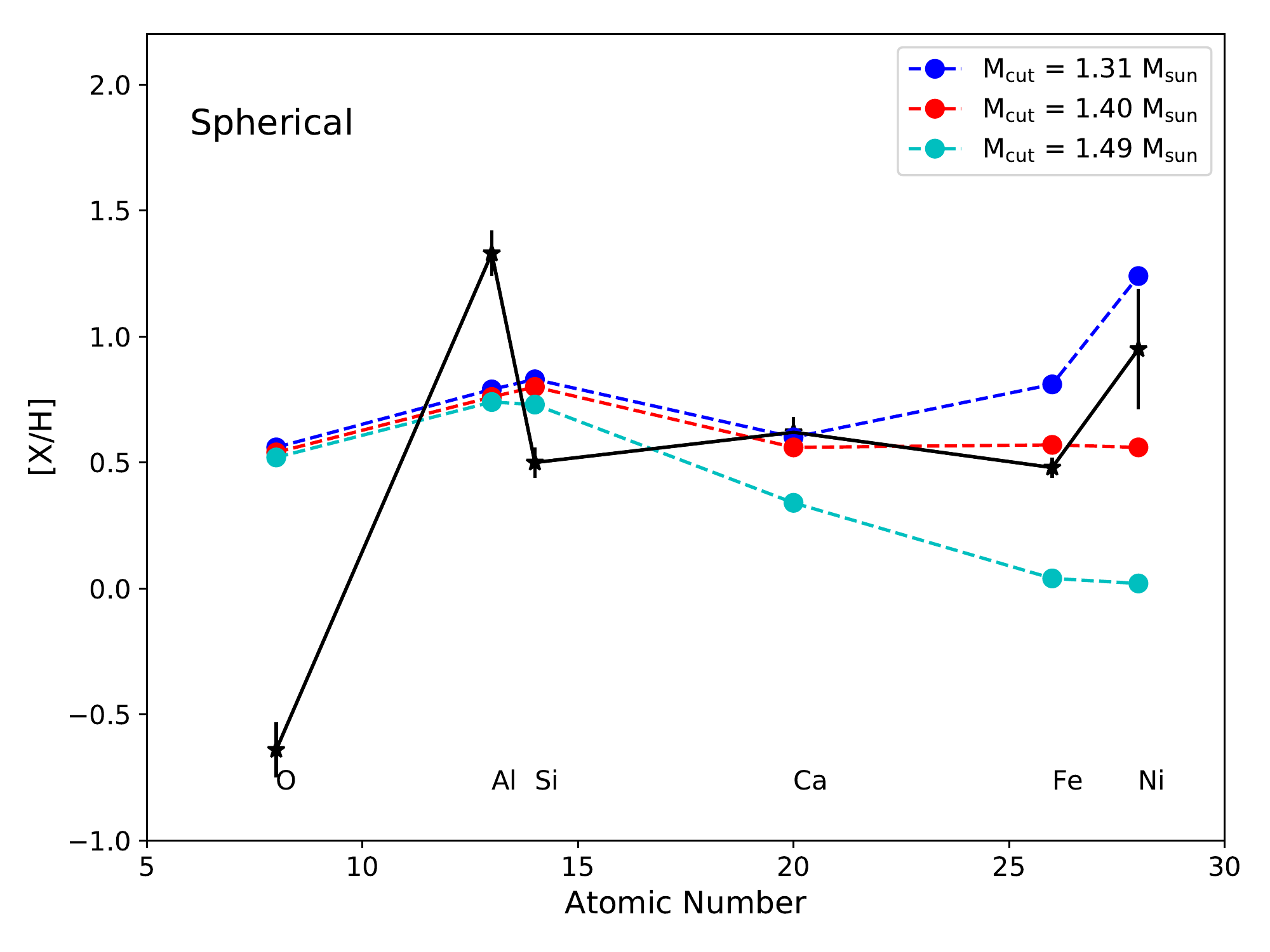}\label{fig:SNa}}

\subfloat[Aspherical symmetric core-collapsed SN explosion.]{\includegraphics[width=0.95\linewidth,angle=0]{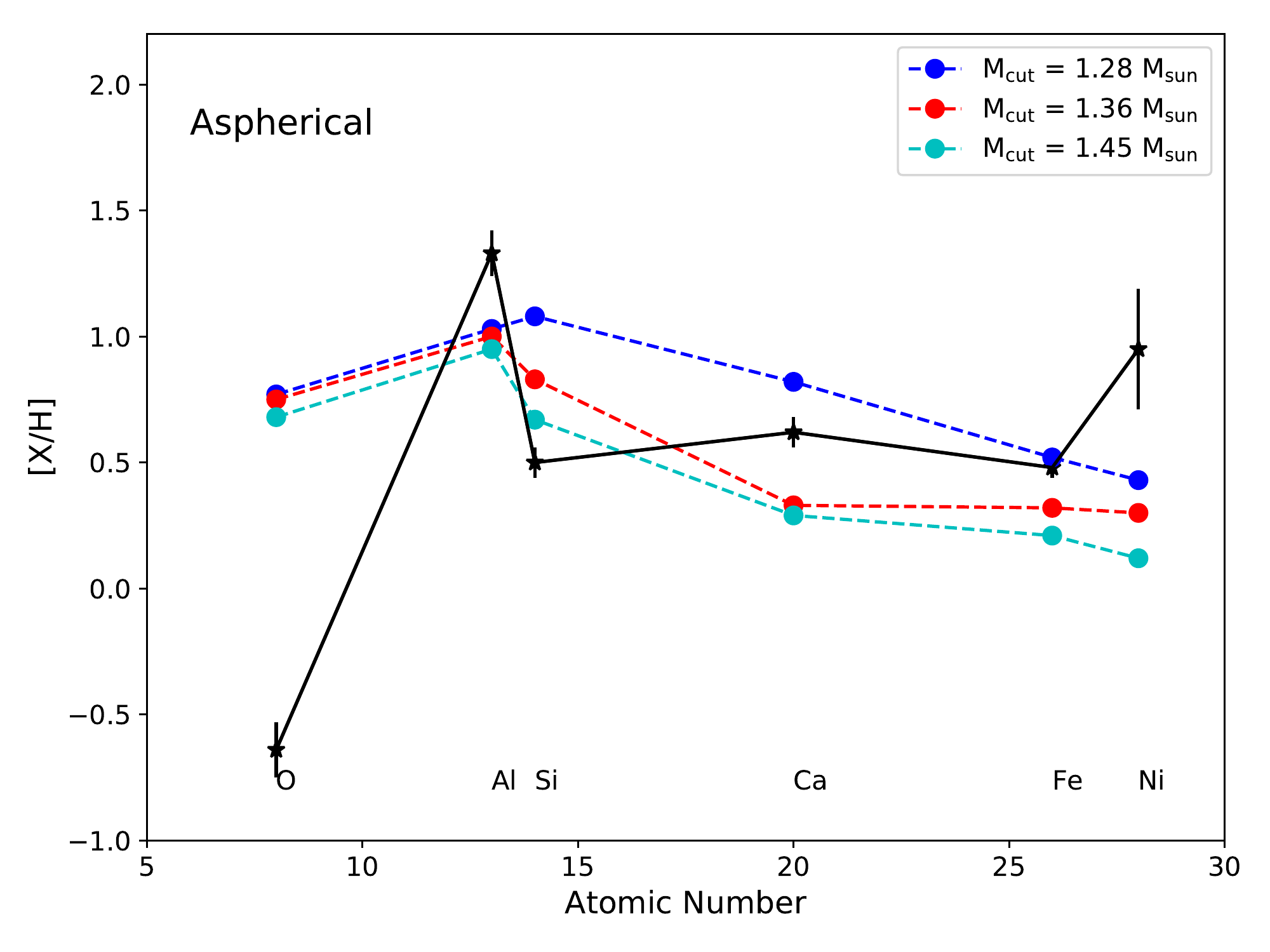}\label{fig:SNb}}

\subfloat[Aspherical symmetric core-collapsed SN explosion with lateral mixing]{\includegraphics[width=0.95\linewidth,angle=0]{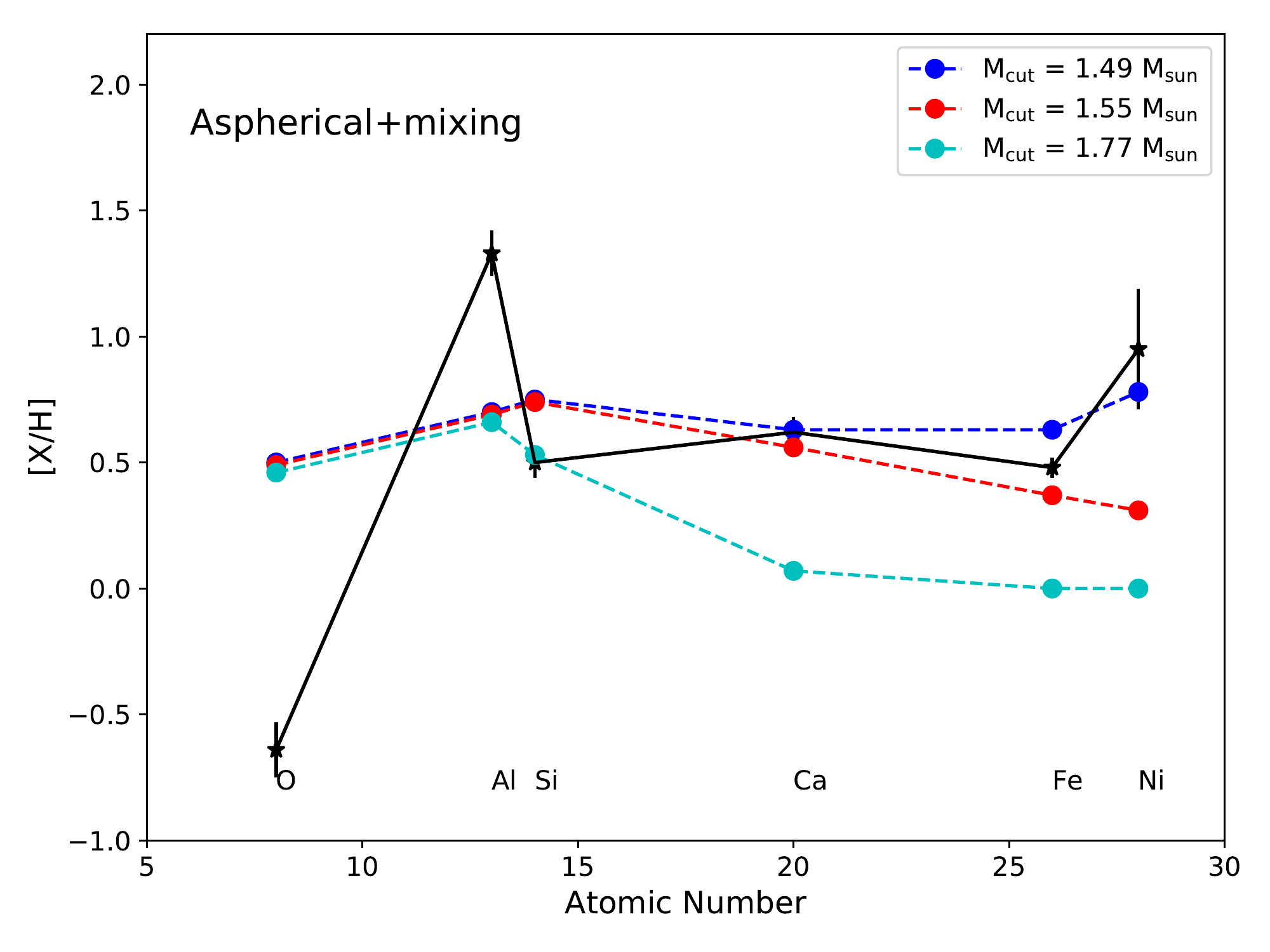}\label{fig:SNc}}
\caption{Expected abundances in the secondary atmosphere due to the pollution caused by (a) a spherically symmetric core-collapsed SN explosion model (top panel), (b) a non-spherically symmetric (or aspherical) SN explosion model (middle panel) and (c) an aspherical SN explosion (bottom panel) with complete lateral mixing (see text and also Table\,\ref{table:SN}). The models are computed for different mass-cuts, $M_{\rm cut}$, assuming a progenitor mass of $M_{\rm He} = 4.0$\Msun, a secondary star initial mass of $M_{2,i} = 1.3$\Msun, and fixed values for $f_{\rm cap}$ = 0.30, 0.50 and 0.40. In all cases an orbital separation of $a_{c,i} = 3$\Rsun\ and a kinetic energy of $E_K = 1 \times 10^{51}$ erg is assumed. The black stars show our measurements for \target.}
\label{fig:SN}
\end{figure}

\begin{table*}
\centering
\caption{Expected abundances from supernova explosion models in \target.}
\begin{tabular}{lcccccccccccc}
\hline
\hline
 & & & & \multicolumn{3}{c}{Spherical$^{\rm a}$} & \multicolumn{3}{c}{Aspherical$^{\rm b}$} & \multicolumn{3}{c}{Aspherical*$^{\rm c}$} \\
\cline{5-7}
\cline{8-10}
\cline{11-13}
& & & & \multicolumn{3}{c}{$m_{\rm cut}$[\Msun]} & \multicolumn{3}{c}{$m_{\rm cut}$[\Msun]} & \multicolumn{3}{c}{$m_{\rm cut}$[\Msun]} \\
\hline
Element & A(X)$_{\odot}^{\rm d}$ & [X/H] & $\delta{\rm [X/H]}$ & 1.31 & 1.40 & 1.49 & 1.28 & 1.36 & 1.45 & 1.49 & 1.55 & 1.77 \\
\hline
O  & 8.74 &  $-$0.64  & 0.11 & 0.56 & 0.54 & 0.52 & 0.77 & 0.75 & 0.68 & 0.50 & 0.49 & 0.46 \\ 
Al & 6.47 &   1.35  & 0.09 & 0.79 & 0.76 & 0.74 & 1.03 & 1.00 & 0.95 & 0.70 & 0.69 & 0.66 \\
Si & 7.55 &   0.50  & 0.06 & 0.83 & 0.80 & 0.73 & 1.08 & 0.83 & 0.67 & 0.75 & 0.74 & 0.53 \\
Ca & 6.36 &   0.62  & 0.06 & 0.60 & 0.56 & 0.34 & 0.82 & 0.33 & 0.29 & 0.63 & 0.56 & 0.07 \\ 
Fe & 7.50 &   0.48  & 0.04 & 0.81 & 0.57 & 0.04 & 0.52 & 0.32 & 0.21 & 0.63 & 0.37 & 0.00 \\ 
Ni & 6.25 &   0.97  & 0.23 & 1.24 & 0.56 & 0.02 & 0.43 & 0.30 & 0.12 & 0.78 & 0.31 & 0.00 \\ 
\hline
$Q^{\rm e}$ &  &     &  & 0.30 & 0.25 & 1.19 & 0.65 & 0.64 & 0.90 & 0.11 & 0.48 & 1.44 \\ 
\hline
\end{tabular}
\begin{tablenotes}
  \item $^{\rm a}$ Spherically symmetric explosion (Spherical). 
  \item $^{\rm b}$ Non-spherically symmetric explosion (Aspherical). 
  \item $^{\rm c}$ Non-spherically symmetric explosion with complete lateral mixing (Aspherical*). 
  \item $^{\rm d}$ Solar abundances are taken from \citet{Grevesse96} and \citet{Ecuvillon06} for O.
  \item $^{\rm e}$ O and Al have not been included.
\end{tablenotes}
\label{table:SN}
\end{table*}

\subsection{Spherical Explosion}
\label{sec:SN_sph}

Following the arguments outlined in \citet{V404Cyg}, we can estimate the maximum ejected mass in the SN explosion that could be captured by the secondary star. \citet{Shahbaz19} estimate the current neutron star and secondary star mass to be  $M_{{\rm NS},f} \sim 1.76$\Msun\ and $M_{2,f} \sim 0.24$\Msun, respectively. Furthermore, the neutron star is expected to have accreted material during the  binary evolution, and so we adopt a canonical neutron star initial mass of 
$M_{{\rm NS},i}\sim$1.4\Msun.

The binary system will survive a spherical SN explosion if the following condition for the ejected mass is satisfied, $\Delta M = M_{\rm He} - M_{{\rm NS},i} \leq (M_{\rm He} + M_{2,i})/2$,
where $M_{\rm He}$ is the helium core mass of the progenitor star before the SN. For the sake of the argument, if we adopt  $M_{\rm He} \sim 4$\Msun, an ejected mass of $\sim2.6$\Msun\ implies an initial mass for the secondary star of $M_{2,i}\geq1.2$\Msun. Therefore, in what follows we assume a secondary star mass of $M_{2,i}\sim1.3$\Msun. Furthermore, using the expressions given by \citet{Zwart97} one can infer mass of the progenitor star, $M_{\rm 1,i} \sim 16.7$\Msun, and the radius of the helium core $R_{\rm He} \sim 0.9$\Rsun. The current orbital period of \target\ measured from pulsar timing \citep{Archibald13} is $P_{\rm orb} \sim 0.198$\,d ($\sim 4.754$\,hr). 
This, together with the mass of the neutron star, the secondary star and Kepler’s third law, gives a current orbital separation of $a_{c,f} \sim$1.79\Rsun. However, during the binary evolution of the system the secondary must have experienced mass and angular momentum losses leading to a decrease in the orbital separation to its present configuration. We therefore assume the post-SN orbital separation after tidal circularisation of the orbit to be $a_{c,i} \sim$ 3\Rsun.


Assuming a pre-SN circular orbit and an instantaneous, spherically symmetric ejection one can estimate the orbital separation just before the SN, $a_{0}$, and after tidal  circularisation, $a_{c,i}$, of the post-SN eccentric orbit using the relation given by \citet{Heuvel84}: $a_{0} = a_{c,i} \mu_{f}$, where $\mu_{f} = (M_{\rm NS} + M_{2,i}) / (M_{\rm He} + M_{2,i})$.  We adopt a secondary initial mass of $M_{2,i} \sim 1.3$\Msun, which implies a pre-SN orbital separation of $a_{0}\sim 1.5$\Rsun, which is larger than the estimated radius of the He core. The secondary star at the time of the SN explosion (at about 7\,Myr)  would still be in the pre-main sequence evolutionary stage and so will have a radius of $R_{2,i} \sim1.5$\Rsun, estimated from the pre-main sequence  \textsc{parsec}\footnote{\textsc{parsec} isochrones are available at http://stev.oapd.inaf.it/cgi-bin/cmd } isochrone \citep[see e.g.][]{Bressan2012}. Part of the matter ejected in the SN explosion is expected to be captured by the  secondary star, thus polluting its atmosphere with SN nucleosynthesis products. The fraction of the matter ejected in the direction of the secondary star that is finally captured can be estimated as  $m_{\rm cap} = \Delta M (\pi R_{2,i} /4 \pi a_{0}^2) f_{\rm cap}$, where $f_{\rm cap}$ is the fraction of mass ejected within the solid angle subtended by the secondary star, that is eventually captured. We assume that the captured mass, $m_{\rm cap}$ (heavier matter compared to the stellar material of the secondary star) is completely mixed with  the rest of the companion star.

We compute the expected abundances in the secondary star after the pollution from a SN explosion as in \citep{A0620,J1655}. We firstly examine a spherically symmetric core-collapse SN of a 4 \Msun\ He core \citep{Maeda02,CenX4} with a  explosion energy of $E_K = 10^{51}$\,erg (see Table\,\ref{table:SN}). We adopt a companion star with a mass of $M_{2,i} \sim$ 1.3\Msun with initial Solar abundances and an orbital separation of $a_{c,i} = 3$\Rsun. We find that we require a capture efficiency of $f_{\rm cap}$ = 0.10--0.30 in order to reproduce the  observed Fe abundance increase from the initial Solar value, which depends on the mass cut, $m_{\rm cut}$. 
The mass cut can be understood as the initial mass of the compact object. For the case of neutron stars, we assume there is no fallback material, so the mass cut reflects the initial mass of the neutron star, and the ejected mass is just the difference between the mass of the He core and the mass cut. 
In Fig.\,\ref{fig:SNa} we compare the observed abundances in \target\ with expected abundances from three different spherical models. We assume a capture efficiency of $f_{\rm cap}$ = 0.30 and different mass cut values (see Table\,\ref{table:SN}). In order to quantify the comparison between  the  observed and model  abundances we determine the variance, $Q$.

The mass captured in these three models are in the range 0.16--0.20\Msun. Fe-peak elements such as Fe and Ni are formed in the inner layers of the SN explosion, and therefore the lower the mass cut, the more Fe and Ni products in the ejecta that can eventually enrich the atmosphere of the secondary star (see Fig.\,\ref{fig:SN}). $\alpha$-elements such as Si and Ca are less sensitive to changes in this mass cut range  1.31--1.49\Msun. Si appears to be about 0.4\,dex more abundant for all mass cuts, while Ca seems to be consistent for mass cuts below 1.4\Msun\ in these models with respect to the observed abundances. O and Al remain mostly independent of the adopted mass cut as these elements form in the outer layers of the SN explosion and, in all models, O is too high and Al is too low with respect to the observed abundances in \target.

\subsection{Non-spherical (aspherical) explosion}
\label{sec:SN_nsph}

The neutron star  in \target\ could have been formed in a non-spherically symmetric explosion. Chemical abundance studies in two other neutron star X-ray binaries,  Cen\,X--4 \citep{CenX4} and  Cyg\,X--2 \citep{CygX2}, arrive at different conclusions.  The case of Cen\,X--4  favors a spherical explosion, whereas the observed abundances in Cyg\,X--2 seems to be better reproduced using aspherical explosion models.  The general scenario extracted from the analysis of the spectra of SN is that neutron stars can form via normal less-aspherical SN explosions 
($E_K = 10^{51}$erg) from relatively low mass primary stars ($M_1 \leq 20$ \Msun), as suggested by \citet{Umeda2003}. A more massive primary star ($M_1 \sim 40$\Msun) can explode as an aspherical hypernova (HN) with a kinetic energy of $E_K = (10-30)\times 10^{51}$\,erg, 
leaving behind a black hole, as proposed to explain the enhanced observed abundances of  some $\alpha$-elements in the black hole X-ray binary Nova\,Sco\,1994 \citep[see e.g.][]{Israelian1999,Brown2000,Podsiadlowski2002}. However, \citet{J1655} performed a detailed chemical analysis of Nova\,Sco\,1994, including additional elements such as Al and Ca, which are formed in the outer and inner layers of the explosion respectively. They concluded that a spherically symmetric and more energetic HN model is favored in this system.  

Large systemic velocities are observed in neutron star X-ray binaries, $\gamma \sim$ +190\kms\ and $\gamma \sim$ $-$210\kms\ for the Cen\,X--4 \citep{Casares07} and Cyg\,X--2 \citep{Casares10}, respectively. The large systemic velocities can be interpreted as a kick in the formation of the compact object due to an asymmetric SN/HN explosion. In contrast \target\ has a small systemic velocity $\gamma \sim$ 5.2\kms\ \citep{Shahbaz19}. Using the parallax and proper motions from Gaia eDR3 \citep{Gaia20} with the observed systemic radial velocity we can determine the 3D velocity of the system (i.e. the velocity relative to expected motion in the Galaxy). We use the method  outlined in \citet{Reid09} to transform the observed parameters  into heliocentric space velocities, and then remove Solar motion as well as Galactic rotation and find a peculiar velocity of 131\kms.

We can compare the peculiar velocity to the expected impulse velocity due  to symmetric mass ejection during a SN explosion. Assuming a 4.0\Msun\ He core and a secondary star of 1.3\Msun\ which produces a bound system with a 1.4\Msun\ neutron star and a post-SN orbital separation of $a_{c,i} \sim$ 3\Rsun, provides an impulse  \citep[Blaauw-Boersma kick;][]{Blaauw61,Boersma61} velocity of $v_{\rm sys} \sim$ 192\kms\ \citep{Nelemans99,Brown01}. Assuming a He core mass of 3.0\Msun\ and a secondary mass of 1.3\Msun\  produces an impulse velocity of $v_{\rm sys} \sim 118$\kms. These observed velocities do not appear to support a significant mass ejection in the SN explosion and suggest that an additional natal kick from an asymmetric SN may not be required. However, aspherical explosions can produce chemical abundance anomalies in the ejecta that cannot be produced in spherically symmetric explosion models. Therefore, in the following we also examine the possible contamination of the atmosphere of the secondary star from a non-spherically symmetric (aspherical) SN. 

A non-spherical SN explosion can produce chemical inhomogeneities in the ejecta which are dependent on direction. If the collimated jet of an aspherical SN explosion is perpendicular to the binary orbital plane, elements such as Ti, Fe, and Ni are ejected mainly in the jet direction, while O, Mg, Al, Si, and S are preferentially ejected near the equatorial plane of the helium star \citep{Maeda02} and thus expected to be enhanced in the atmosphere of the secondary star. 

In Fig.\,\ref{fig:SNb} we show the expected abundances in the secondary's atmospheres from an aspherical explosion of a 4.0\Msun He core. Fe and Ni are preferentially ejected perpendicular in the orbital plane only with a mass cut below 1.3\Msun and with a capture efficiency of $f_{\rm cap}$ = 0.50, which implies a significant fraction of mass (0.30--0.36\Msun) is captured by the secondary. Even with these parameters the amount of Ni in the SN model is about 0.5\,dex lower than the observed abundance. Si, Ca, Fe and Ni are are highly dependent on mass cut in the range 1.28--1.45\Msun. In particular Ca is only highly enhanced for the  lowest mass cut at 1.28\Msun. O and Al are mainly ejected in the equatorial plane, and thus only depend on the capture efficiency, and they behave similar to the spherical model (Fig.\,\ref{fig:SNa}) with respect to the observed abundances.

Finally, in  Fig.\,\ref{fig:SNc}, we also consider an aspherical explosion model with the extreme assumption of complete lateral mixing  \citep[see][for further details]{Maeda02,Podsiadlowski2002}, where the ejected matter is completely mixed with each velocity bin \citep{J1655}. This model requires $f_{\rm cap}$ = 0.40 with a captured  mass of 0.19--0.22\Msun. As in the case of the spherical SN explosion, the expected abundances of Si, Ca, Fe and Ni show a dependence with mass cut in the range 1.49--1.67\Msun. Qualitatively, the spherical model with $m_{\rm cut}$ = 1.31\Msun\ and the aspherical model with $m_{\rm cut}$ = 1.49\Msun\  agree better with the observations. However, as with all the other models, we cannot reconcile the O and Al observed abundances. 

\citet{Nomoto10} argue that more energetic HNe are produced by more massive progenitors of compact objects (likely black holes) which may be more asymmetric than normal SNe that show different levels of kinetic energies \citep{Maeda06}. We have not been able to favor the spherical explosion over the aspherical in the case of \target. The high element abundances derived in \target, in particular, the high content in Fe-peak elements support the formation of a neutron star in a SN. This new Fe measurement in this MSP contributes to the detection of Fe in other neutron star X-ray binaries where also a high Fe abundance has been reported and supports the idea that the formation of neutron star in normal less massive SNe ejects a significant fraction of $^{56}$Ni. However, \citet{Nomoto10} also argue that more energetic HNe that may led to the formation of black holes are expected to eject larger quantities of $^{56}$Ni than less massive progenitors of neutron stars that explode with lower kinetic energies and presumably smaller amounts of $^{56}$Ni.

\subsubsection{Other chemical elements}

In the following sections we discuss also the evolution of CNO and Li in the atmosphere of the secondary star. C and N are produced in the outer layers of both spherical and aspherical SN explosions. The expected abundances of C and N follow the same behaviour as O in the models with lower levels of production at [C/H] = 0.40--0.60 and [N/H] = 0.10--0.15. Unfortunately, we do not have C and N features to measure abundances in \target\ in the spectral range available in this work. Li is not present in these SN explosion models. The mass fraction of Li in the ejecta of the models evaluated in this work is negligible. The production of Li in core-collapse SN has been studied by \citet{Woosley95}, suggesting $^7$Li production as a consequence of a sufficiently high neutrino irradiation in the SN model. However, \citet{Prantzos12} argues against a significant role of core-collapse SN in the production of Li in the Galaxy.

\section{Chemical abundance anomalies due to CNO evolution}
\label{sec:cno_evol}

\begin{figure}
\centering
\includegraphics[width=1.0\linewidth,angle=0]{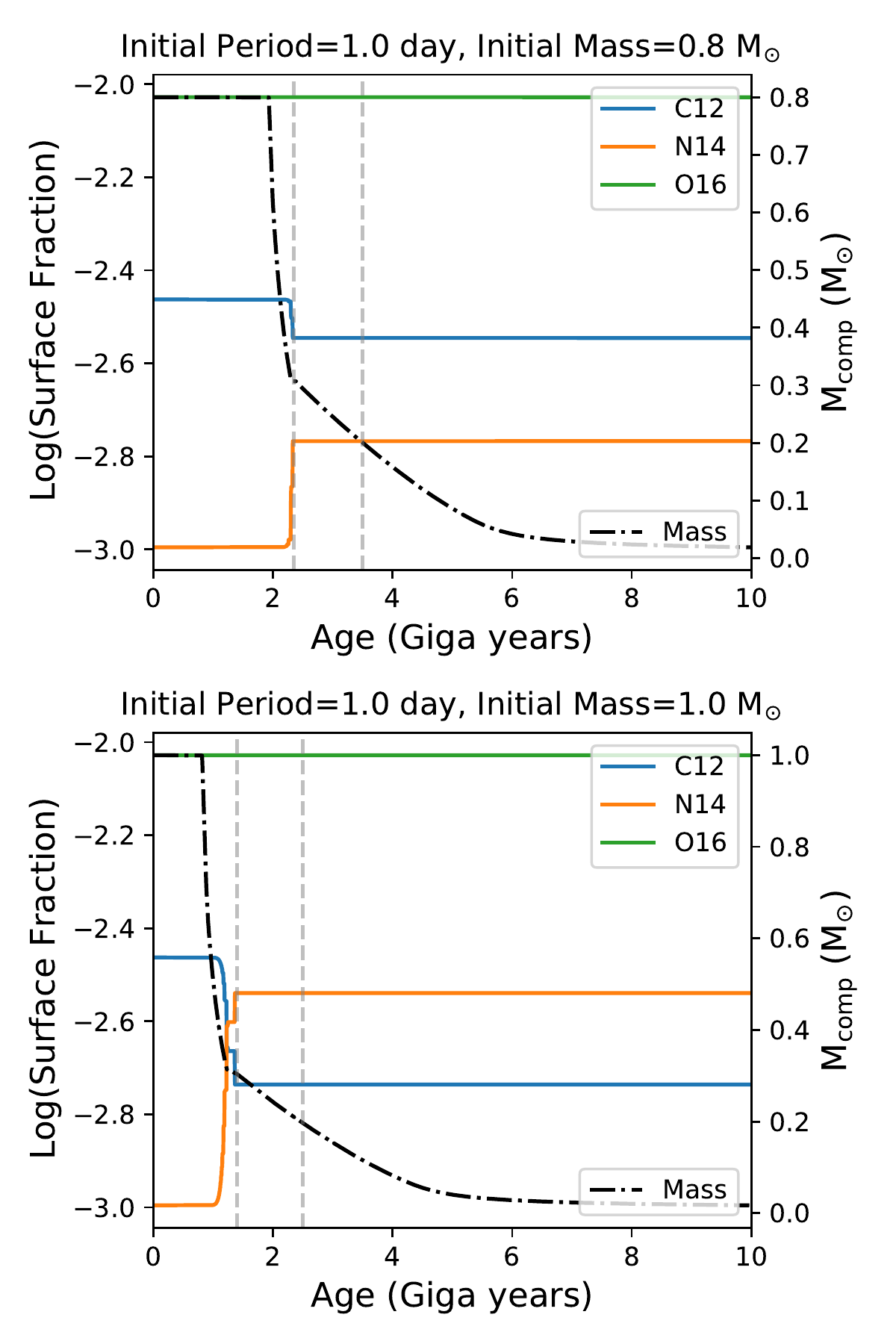}
\caption{The results of \textsc{mesa} simulations using the angular momentum loss of \citet{Chen13}. The initial period of the binary is 1\,d, while the initial masses of the secondary star are 0.8\Msun\ (top plot) and 1.0\Msun\ (bottom plot). The $^{12}$C (blue), $^{14}$N (orange) and $^{16}$O (green) elemental surface fraction are shown as time. The secondary star's mass (black dot-dashed line) remains constants until it comes into contact with the Roche lobe and then decreases very rapidly until it reaches $\sim$ 0.3 \Msun, in which case magnetic braking turns off, and mass transfer slows down. The  C and N surface composition after t $\sim$ 2.2\,Gyr on the left plot ($\sim$ 1.32\,Gyr on the right plot) remains constant because magnetic braking has turned off, even though the companion star continues to loose mass. } 
\label{fig:MESA_plots}
\end{figure}

After the SN explosion, the subsequent evolution is driven by strong mass-loss, most of which is accreted by the neutron star. Over the last number of years, the consensus that has developed in the community is that the efficiency of angular momentum loss  from  the binary, whether driven by magnetic braking or through irradiation feedback, is a key factor in determining whether a binary evolves into a black widow, a redback, or a pulsar and white dwarf binary \citep{Chen13, Jia15, Van19a, Van19b}. This efficiency is inherently linked to the evolutionary status of the companion. As a consequence of mass-loss and nuclear evolution, a convective envelope develops and penetrates to deep layers of the star, which are then mixed, changing the star’s surface chemical composition. If the star’s surface is exposed to material that has undergone partial CNO burning, its composition should show the signature of CNO processing where C depletion (and if significant CNO burning has occurred, also O depletion)  and N enhancement are expected \citep{Clayton83}. As such, the C/O ratio and C/N ratio of a given binary can be used to infer the evolutionary status of the companion star prior to the start of mass transfer \citep{Nelemans10}. This has been successfully applied to differentiate the formation channels for the hydrogen deficient companions of white dwarfs in short orbital period ($<$70 min) systems (e.g. \citealt{Kennedy15}; \citealt{Kupfer16}), where the ratio of N and O relative to C has revealed whether the companion is a low mass helium star or slightly evolved, post-main sequence star. It has also been seen in an X-ray binary system \citep{Ergma01} that there is a decrease in the surface $^{12}$C/$^{13}$C ratio, with a corresponding increase in $^{14}$N and decrease in $^{15}$N, so the $^{14}$N/$^{15}$N ratio increases with respect to the initial value. The $^{16}$O abundance remains unaltered, while that of $^{17}$O increases.

The evolutionary models proposed for redbacks and black widows are also capable of making predictions for the chemical abundances at the stellar surface of the companion. While a full, detailed study of the effects of the initial binary parameters and the strength of the angular momentum loss on the chemical composition of redback companions is beyond the scope of the current work, we have reproduced selected models discussed in \citet{Chen13} and in \citet{Van19b} using Modules for Experiments in Stellar Astrophysics (\textsc{mesa} version 15140; \citealt{MESA1}; \citealt{MESA2}). The C, N and O abundance at the surface appears to be tightly controlled by a combination of the rate of angular momentum loss and the initial conditions of the binary. For inefficient angular momentum loss and where the donor star evolves significantly before the onset of mass transfer, we produce companions which resemble the hydrogen deficient companions of white dwarfs previously mentioned, with no detectable H or C at their surface, very little O, and significant N. This is at odds with the spectrum of the companion of \target, which still shows strong hydrogen features.  For efficient angular momentum loss, we produce companions with overabundant O, slightly overabundant N, no C, and typical amounts of H. 

Fig.\,\ref{fig:MESA_plots} shows the results of two intermediate cases, where we have used the angular momentum loss of \citet{Chen13}. In both, the initial period of the binary is 1\,d, while the initial masses of the donor star  are 0.8 \Msun\ and 1.0 \Msun, respectively. As mass-loss from the secondary star commences, the  $^{14}$N/$^{12}$C ratio quickly changes, reaching a new equilibrium value once the star has become fully convective and angular momentum loss due to magnetic breaking switches off. This shows that we do expect modest  $^{14}$N  enhancement in binaries which produce companions similar to the one in \target, while  $^{16}$O is untouched in these models. A more robust study on the relationship between the chemical composition of the secondary stars in spider binaries and their binary evolution is required to better understand the anomalous abundances seen in \target.

The only MSP where a spectroscopic chemical abundance analysis exists is for PSR\,J1740--5340, which lies in the globular cluster NGC\,6397 \citep{DAmico01a}. PSR\,J1740--5340 is a redback MSP which has a $\sim$0.3\Msun\ secondary star in a 32.5\,hr binary orbit \citep{Ferraro01,Orosz03}. 
There is a complete absence of C and the enhanced N in the atmosphere indicates a composition resulting from the hydrogen-burning CNO cycle. \citet{Ergma03} computed evolutionary models for PSR\,J1740--5340 to predict the surface chemical composition as a function of the secondary mass. For the case of an evolved star which has lost mass, O would have a normal abundance, whereas N would be overabundant and C under-abundant, suggesting that the secondary star in PSR\,J1740--5340 is a low-mass (0.3 \Msun) remnant star of a deeply peeled star 0.8\Msun\ progenitor \citep{Mucciarelli13}. 

In \target\ we find [O/Fe] is under-abundant compared with the Galactic trends of stars in the Solar neighbourhood. Unfortunately, due to our spectral range coverage we cannot measure the C or N abundances in \target. This means that we cannot fully test CNO processing models. However, given that one expects normal O abundance, in contrast to what we observe suggests that underabundant O in \target\ is not due to chemical anomalies associated  with the CNO cycle.

\begin{figure}
\centering
\includegraphics[width=1.0\linewidth,angle=0]{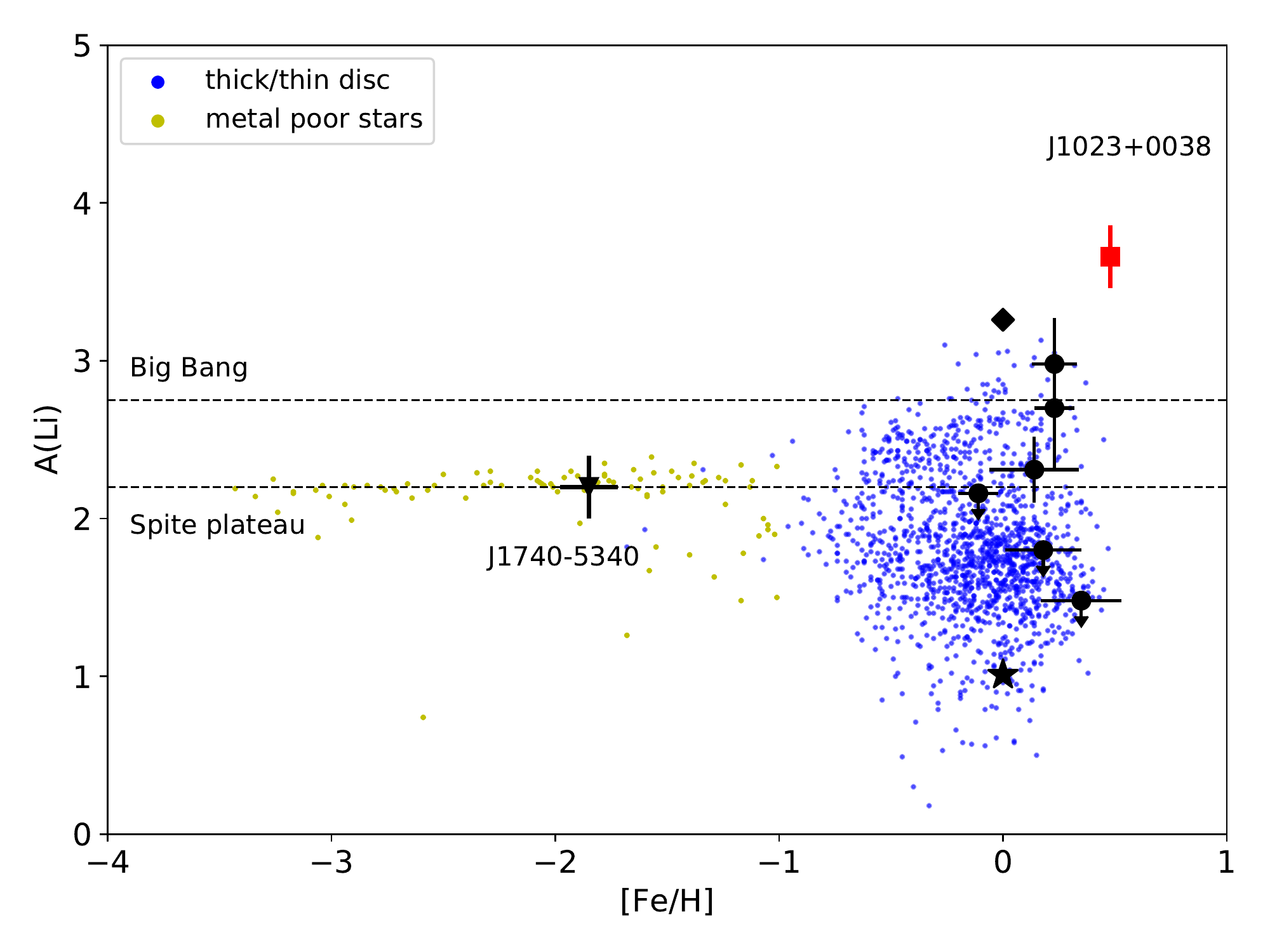}
\caption{Li abundances as a function of [Fe/H]. The yellow and blue dots show the Li abundance of metal poor stars \citep{Melendez10} and the thick/thin disc \citep{Fu18}, respectively. The primordial $^7$Li abundance value A(Li) = 2.75\,dex determined from Standard Big Bang Nucleosynthesis  models \citep{Pitrou18} and the `Spite plateau' at A(Li) $\sim$ 2.2\,dex determined from  metal-poor Population II halo dwarf stars \citep{Spite82} are shown as dashed lines. The value we obtain for \target\ (red square) is also shown along with the value obtained for the black widow pulsar PSR\,J1740--5340 (black filled triangle; \citealt{Sabbi03}) and the values obtained in X-ray binaries (black filled circle; see Table\,\ref{table:Li_XRBs}).
We show the Solar value (black star) of A(Li) = 1.01\,dex \citep{Anders89}, meteoritic value (black diamond) of A(Li) = 3.26\,dex \citep{Asplund09}.} 
\label{fig:FeLi}
\end{figure}

\section{Lithium around compact objects}
\label{sec:Li}

Li is a complex element and can be created in a number of ways such as Big Bang nucleosynthesis, in the cool bottom burning process during the asymptotic giant branch (AGB) phase, or via spallation of CNO nuclei. On the other hand, it is easily destroyed in the stellar interiors. 

The primordial Li abundance is preserved in stars that are old and comparatively cool. Since nuclear-fusion reactions take place in the star's inner hot regions, the composition of the outermost layers of old stars indicates the initial chemical content at the time of  the star's formation and so abundances should be close to the primordial values. The observed `lithium plateau' in the $^7$Li abundance of metal-poor Population II halo dwarf stars at A(Li) $\sim$ 2.2\,dex is well-established \citep{Spite82} and is often interpreted as due to a depletion in $^7$Li from the primordial cosmological value of A(Li) = 2.75\,dex predicted by  Standard  Big Bang  Nucleosynthesis models \citep{Pitrou18} in the course of stellar evolution \citep[e.g.][]{Charbonnel05,Cyburt08}. This discrepancy is generally denominated as the `cosmological lithium' problem. Many studies have been devoted to solve this problem  \citep[see][for a review]{Mathews20}. 

As stars evolve Li is pulled to hot deep layers and gradually destroyed via proton capture ($p,\alpha$) reactions in the hotter stellar regions with a temperature of a few 10$^6$\,K  \citep{Pinsonneault97}. Li at the surface layers is also expected to be depleted via various  processes, such as atomic diffusion \citep{Michaud86}, rotational mixing \citep{Pinsonneault92} or convection overshooting \citep{Xiong09} which can lead to a significant reduction in the Li abundance in the star’s atmosphere as it evolves.

The Li abundance in young  Population I stars is equal to the meteoritic value of A(Li) = 3.26\,dex \citep{Asplund09} which is an indirect  indication of the Li abundance in the interstellar gas-dust medium from which these stars were formed. This is an order of magnitude greater than the Li abundance found in old stars in the halo (the `lithium plateau') which indicates that the Galaxy has undergone a history of Li enrichment since the Big Bang. To reach the high $^7$Li values found in meteorites, which is generally accepted to be the typical initial Li abundance of Population I stars, requires Galactic Li enrichment. Mechanisms of Li enhancement in the Galaxy include cosmic-ray spallation interactions in the interstellar medium \citep{Reeves70}, core-collapse SN, \citep{Woosley90}, novae \citep{Hernanz96} or in evolved low-mass  stars \citep[see][and references within]{Prantzos17, Matteucci20}.

Li is easily ionized (its ionization potential is 5.39\,eV), so Li\,I line is only observed in relatively cold stars with effective temperatures less than 8500\,K corresponding to stars with a spectral type ranging from late A to M.  The analysis of the strong Li\,I resonance line at 6707.8\,\AA\  provides data for the $^7$Li abundance in these stars. In Fig.\,\ref{fig:FeLi} we show the $^7$Li abundance of metal-poor  dwarf stars and the primordial $^7$Li abundance value, as well as the $^7$Li abundance we observe in \target.

The observed Li abundance in neutron star and black hole quiescent X-ray transient binaries is relatively high, but not in excess of the meteoritic value (see Table\,\ref{table:Li_XRBs}). This is surprising as Li is destroyed/depleted in stellar interiors. However, Li can also be preserved. Heating of the secondary star's surface by high energy radiation produced near the compact object  may stabilize the outer layers to convection, so that the light elements in these layers are never driven to the depths at which they are destroyed \citep{Eichler96}. Rotation might also reduce Li suppression mechanisms, in that the tidally locked rotation of the secondary star naturally leads to slower Li destruction rates leading to Li abundances closer to the standard value at formation for Population I stars \citep{Maccarone05}. To some extent this is supported by the measurement of the $^6$Li/$^7$Li  isotopic ratio in Cen\,X--4 \citep{Casares07}. Clearly, the definite proof of Li production would be to determine a Li abundance higher than the cosmic value.

In Fig.\,\ref{fig:FeLi} we show the $^7$Li primordial abundance and the abundance in metal-poor galactic Halo and thin/thick disc stars. We also show the $^7$Li abundance measured in \target\ 
with A(Li) = 3.66 $\pm$ 0.20\,dex which is greater than the meteoritic value and what is observed in young Population I stars at the 2$\sigma$ level, providing evidence for Li enhancement in the atmosphere of the secondary star. It should be  noted that enhanced Li is also observed in the redback pulsar PSR\,J1740--5340 containing a subgiant secondary star with A(Li) = 2.2 $\pm$ 0.2\,dex, which is higher than what is expected for evolved subgiant stars A(Li) $\sim$ 1.5\,dex \citep[see][and references within]{Sabbi03}. As suggested by the authors, the most plausible explanation is fresh Li production due to nuclear reactions occurring on the stellar surface induced by the cosmic-rays produced by the pulsar. In the following sections we outline plausible explanations for the Li enhancement in \target.

\section{Li enhancement}
\label{sec:enhancement}

\subsection{Spallation via neutrons }
\label{sec:neutrons}

As outlined in various papers, the conditions around compact objects are ideal for Li production via the acceleration of relativistic particles and spallation in the inner accretion flow or in the stellar atmosphere \citep{Martin95,Yi97,Guessoum99,Fujimoto09}. The quiescent black hole and neutron star X-ray binaries have hot advection-dominated accretion flows (ADAFs; \citealt{Narayan95b,Narayan96, Menou99}). The ion temperature around black holes and neutron stars is very similar and can be predicted by the basic assumptions set in the ADAF in which the ions are first heated by the viscous processes, and then a fraction of this heat energy which is stored in the ions is transferred to  the  electrons  via  Coulomb collision.  Given the low  gas density in the ADAF, the  Coulomb  collision is not efficient and so the amount of heat energy transferred to the electrons is small. Consequently, the ions are kept at a relatively higher temperature, which is close to the virial temperature,  $\rm \sim 10^{ \rm 12}$\,K near the compact object. As a consequence, nuclei with energies $>$10\,MeV per nucleon are abundant and Li production via $\alpha - \alpha$ spallation of CNO nuclei by neutrons is possible inside the ADAF region \citep{Yi97}. Li production is also possible on the surface of the secondary star \citep{Guessoum99, Fujimoto08}. Given that neutrons are relatively easy to produce in ADAFs either via $\rm p-\alpha$ or $\alpha-\alpha$ reactions and a large fraction are produced with sufficient energy that they do not interact with nuclei through the Coulomb interactions and so escape from the gravitational potential of the compact object  \citep{Guessoum90}. The neutrons ejected from the hot ADAF are then intercepted by the secondary star and interact with CNO nuclei through spallation to produce Li on the star's surface \citep{Guessoum99}. Indeed, model predictions are in good agreement with the observed values in quiescent X-ray binaries \citep{Fujimoto08}.

In binary MSPs, in principle a  hot, geometrically thick accretion flow (beyond the light cylinder radius, $R_{\rm lc}$) with low ($\rm \sim 10^{-4}$) Eddington accretion rates exist similar to the quiescent X-ray binaries. The pulsar magnetosphere remains active in the disc-state and the accretion disc is truncated at a few light cylinder radii ($R_{\rm lc} \sim$ 80\,km) away from the pulsar  \citep{Papitto19,Veledina19}. Given that the pulsar wind is active in gamma-rays and also irradiates the disc, the ion temperature of inner accretion flow is an unknown quantity. However, \citet{Qiao21} determine the emergent spectrum arising from  an ADAF region around the weakly magnetized neutron star, considering the radiative coupling between the soft photons from the surface of the neutron star and the ADAF. The ion temperature at the inner edge of the disc (at a few light cylinder radii away from the pulsar) is $\sim$ 14\,MeV, which is sufficient to produce Li via CNO nuclei by neutrons inside the ADAF region in a similar way as in X-ray binaries.  A fraction of Li produced in an ADAF \citep{Yi97} could be expelled via magneto-centrifugal propeller effects and transferred to the secondary star enhancing the Li abundance \citep{Fujimoto09}. Also, neutrons expelled from the ADAF and intercepted by the secondary star can interact with CNO nuclei through spallation to produce Li on the star's surface \citep{Guessoum99,Fujimoto08}.

\subsection{Spallation via high energy radiation }
\label{sec:radiation}

The detection of pulsed gamma-ray emission from a large number of MSPs by Fermi Large Area Telescope  \citep{Abdo13} imply copious pair production. In MSPs, protons and electrons are initially extracted from the neutron star surface by the intense rotation-induced electric field and later transformed into electron-positron pairs through electromagnetic cascading \citep{Sturrock71} in the MSP magnetosphere \citep{Venter09,Johnson14}. Some of these pairs are advected into the relativistic pulsar wind, powered by the pulsar's rotational energy and end up as part of the highly relativistic magnetized wind emerging from the pulsar. The collision between the  highly relativistic pulsar wind with the mass outflow from the secondary star produces an intrabinary shock structure \citep{Romani16}. The wind is compressed by the shock and so the wind particles are accelerated to higher energies.  Non-thermal X-rays are produced from the accelerated particles in the shock region \citep{Arons93,Bogdanov11}.  Pair particles energized in the shocked pulsar wind are accelerated to mildly relativistic velocities and beam synchrotron radiation in a hollow cone pattern which is observed in the X-ray band. X-ray emission modulated at the orbital periods have been observed in the black widow systems PSR\,B1957+20 \citep{Huang12}, PSR\,J2215+5135, and PSR\,J2256–1024 \citep{Gentile14} as well as in \target\ \citep{Bogdanov11}

If sufficiently energetic, the intrabinary shock particles can  produce gamma-rays via synchrotron or inverse Compton emission \citep{Romani16,Wadiasingh17,Kandel19,Bednarek14}. There is also some evidence of an additional higher GeV emission component arising from inverse Compton scattering of the secondary star's thermal radiation off a cold ultra-relativistic pulsar wind \citep{Wu12}. Although pulsar winds are usually modeled as pair winds,  such winds may also inject ions \citep{Arons03}. Furthermore, the intrabinary shock can also be very efficient at accelerating particles through the Fermi first-order mechanism \citep{Drury83}. In particular, redbacks (in the pulsar-state) and black widow pulsars can have electron-positron pairs reaching energies of 1--10\,TeV \citep{LK21} and protons with energies as high as 10--100\,TeVs \citep{Harding90}. These protons from the intrabinary shock may subsequently interact with the companion star, and eventually lead to spallation of CNO nuclei producing Li as well. The intrabinary shock can also give rise to ion acceleration, where ions escaping from the pulsar polar caps as the result of thermionic emission \citep{Arons92} are accelerated to high energies by the electric potential induced by pulsar rotation and then accelerated by the intrabinary shock. The interaction of these high energy ions with the CNO nuclei in the secondary star's atmosphere via spallation can also result in an increase, but not significant amount, in the Li abundance \citep{Luo98}.

In the accretion-powered disc-state  the pulsar wind is maximal at the equatorial plane and directly interacts with the accretion flow inner boundary to produce gamma-rays \citep{Veledina19}. They can also accelerate protons to energies well above 1\,TeV  \citep{Papitto14}, which could subsequently lead to both very high energy gamma-rays and neutrinos \citep{Eichler78}. In principle, the high energy gamma-rays in MSPs can interact with the secondary star's atmosphere to produce light elements \citep{Eichler96}. High energy gamma-rays or pairs impinging the secondary star's surface cascade via bremsstrahlung pair production cycles, have photon energies above $\sim$15\,MeV sufficient for CNO nuclei spallation to occur, leading to Li enrichment in the secondary star's atmosphere. For particles such as cosmic-rays, Li production is also possible via spallation processes in which CNO-enriched low energy cosmic-rays accelerated in shocks interact with the ambient interstellar medium to produce Li via spallation (p, $\alpha$ + C, N, O $\rightarrow$ $^{6,7}$Li) or fusion ($\alpha$+$\alpha$ $\rightarrow$ $^{6,7}$Li; \citealt{Reeves70,Meneguzzi71}). 

\begin{figure}
\centering
\includegraphics[width=1.0\linewidth,angle=0]{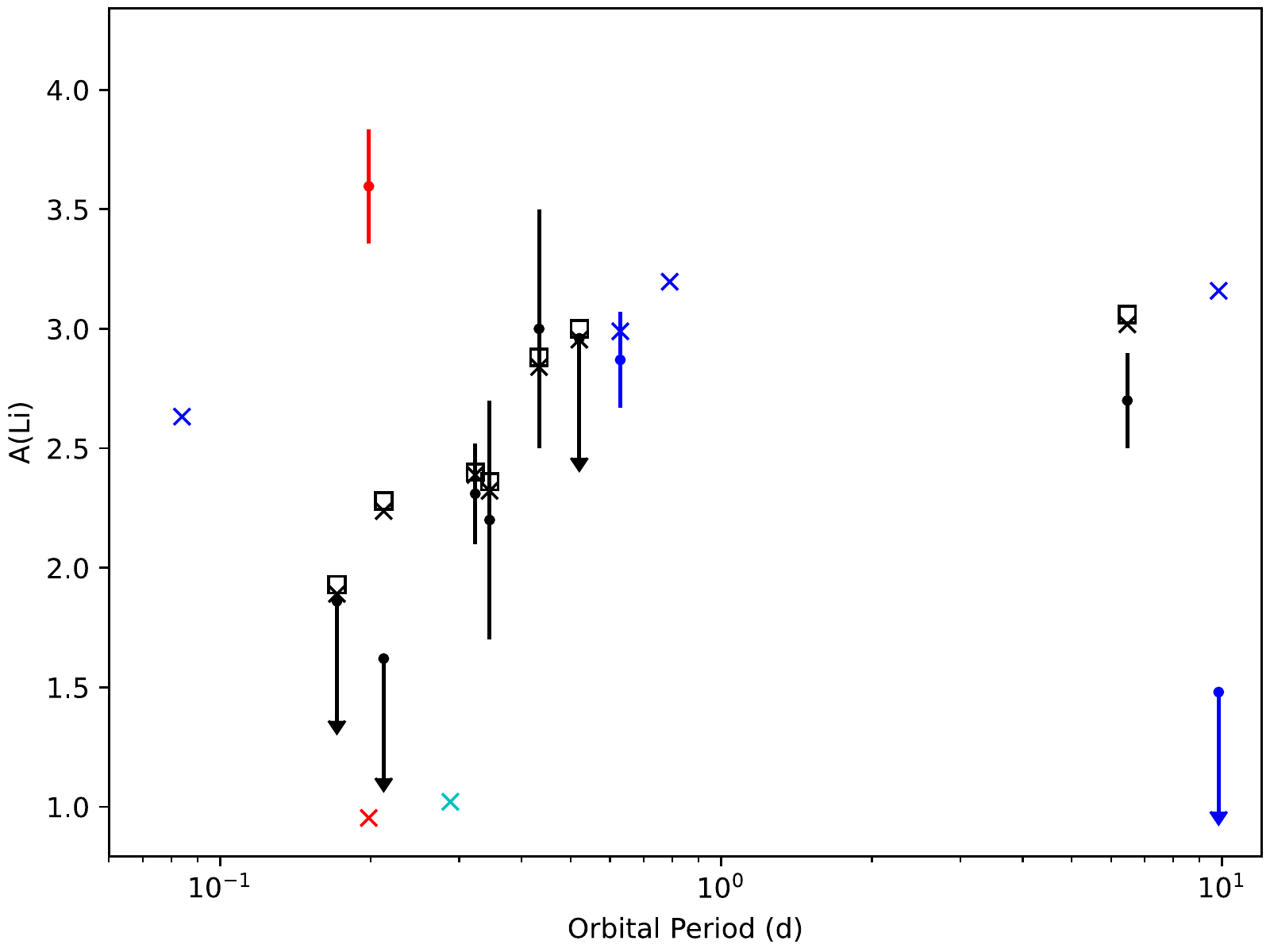}
\caption{Li abundance due to neutron spallation with black hole and neutron star binaries in black and blue, respectively. \target\ is marked in red, and the other Galactic field tMSP PSR\,J1227$-$4853 in cyan. Dots are observed values with their corresponding error bars or upper limits. Open squares are the predicted values based on the model from \citet{Fujimoto09}, with data sourced from this paper for black hole binaries while for neutron star systems they are taken from \citet{Casares07} and \citet{Suarez-Andres15}.
}
\label{fig:spallation}
\end{figure}

\section{Discussion}
\label{sec:discussion}


\subsection{Spallation via neutrons}
\label{sec:discussion:neutrons}

We consider the possibility that the unusually high Li abundance in \target\ is the result of spallation of CNO atoms in the upper layer of the secondary star by high-energy neutrons produced in the accretion disc as suggested in quiescent X-ray binaries \citep{Fujimoto08}. In the following we use the methods outlined in  \citet{Fujimoto08}. Assuming a mass transfer rate from the secondary star of $4 \times 10^{13}$\,g\,s$^{-1}$ \citep{Campana16} which corresponds to a fraction $1.6 \times 10^{-4}$ of the Eddington rate, we estimate the neutron ejection rate from the accretion disk to be $2 \times 10^{31}$\,s$^{-1}$. Using this neutron rate we can calculate the mass layer of the companion exposed to spallation and determine the production rate of Li for the system parameters of \target. Assuming that the main source of Li depletion is the mass transfer from the secondary star's surface to the accretion disc, we calculate an equilibrium Li abundance of A(Li) = 0.95\,dex, which is over 1000 times lower than the observed number fraction. The time-scale for Li enhancement produced from neutron spallation is then $\sim$194\,yr. In Fig.\,\ref{fig:spallation} we show the predicted and observed Li abundance for \target. For comparison we  show the values for the quiescent neutron star and black hole X-ray binaries \citep{Fujimoto08}.  We also show predictions for other MSPs. The main difference between the quiescent neutron star X-ray binaries and MSPs is the much lower accretion rate in the latter, which results in a lower neutron ejection rate  and hence lower Li number fraction.


It should be noted that the parameters that can be modified to reconcile the predicted equilibrium abundance with observations are limited. Quantities such as cross-sections, orbital separation and companion size are only expected to vary by order unity. The equilibrium abundance varies directly with both the mass transfer rate and CNO number fraction, and inversely proportionally with the hydrogen number fraction. Despite the fact that the secondary star should overall be hydrogen-poor and CNO-rich, their linear dependence and the presence of  hydrogen lines in the spectrum implies that the photosphere must retain a somewhat normal hydrogen fraction. It is not feasible for a combination of the hydrogen and CNO number densities to yield the observed Li abundance. Likewise, increasing the mass transfer rate to match the observed Li abundance would bring it to the Eddington rate, which is in direct tension with the low X-ray luminosity of the system. It therefore appears that spallation by neutrons emitted by the accretion disc is not a viable scenario to enrich the secondary star's surface with Li in \target.

\subsection{Spallation via protons}
\label{sec:discussion:protons}

An alternative Li production mechanism is through the spallation of CNO nuclei by protons. Such protons can be accelerated through the pulsar wind and efficiently interact with the companion star's atmosphere. 
Also, protons arising from the companion star's wind can also be accelerated in the intrabinary shock \citep{Harding90}. 
While precise details of the exact energy distribution and number density of protons in the pulsar wind is still an open question, it appears realistic to assume that acceleration to Lorentz factors of a few, if not higher, and proton densities of order $n_{\rm GJ}$, the fiducial Goldreich-Julian density \citep{Goldreich69}. Recent particle-in-cell simulations have for instance demonstrated the proton outflow from the pulsar wind is largely equatorial in structure, mostly independent of the pair production efficiency and has a current density approximately decreasing with the square of the distance \citep{Guepin2020}.

Under these assumption, we can estimate the Li abundance using a similar procedure as the one employed in Section\,\ref{sec:discussion:neutrons}, but adopting p-p scattering as the main opacity source controlling the mass of the atmosphere exposed to proton bombardment and using the appropriate cross section for the proton spallation on CNO nuclei. For the p-p cross-section we adopt an approximate average of $\sigma_{\rm pp} \sim 40$\,mbarn in the few tens of MeV to a few GeV range \citep{Zyla20}. Larger than a 100\,MeV, the spallation cross section of p onto CNO (for the combined $^6$Li and $^7$Li isotopes) averages to around 33\,mbarn \citep{Read84}. We calculate a proton ejection rate of $\sim 10^{34}$\,s$^{-1}$ for a typical MSP magnetic field and spin period. 
This is higher than the neutron ejection rate as the neutron production rate in the disc is driven by destruction of He, while proton production comes directly from the pulsar's wind. Taking mass-loss (feeding the disc in the disc-state and the outgassing material creating radio eclipses in the pulsar-state) to be $\sim 10^{-13}$\,M$_\odot$yr$^{-1}$ as the main depletion source (as we did for the neutron spallation case), we determine an equilibrium abundance A(Li) = 4.45\,dex. This value is therefore compatible with the observed measurement without requiring much fine tuning. Furthermore, the enhancement time-scale, 600\,yr, implies that equilibrium quickly settles. Thus we conclude that proton spallation is a viable mechanism.


\subsection{Spallation via gamma-rays}
\label{sec:discussion:gamma}

An alternative option is that Li is produced  via the photospallation mechanism. \citet{Eichler96} studied the spallation process on the surface of the secondary star induced by the gamma-ray photons from a pulsar wind, which becomes significant at energies $\geq$15\,MeV. They found that the photodisintegration of $^{12}$C, even when affecting a relatively small fraction of the available atoms, produced a significant increase of lighter elements abundances, in particular B, Be and Li. The exact amount of light elements kept in the secondary star's surface depends on different parameters such as the evaporation rate and impinging flux due to the pulsar wind, the mixing depth, the size of the secondary star and the age of the system. A full calculation  is beyond the scope of this paper, but we can make some estimates of the expected Li abundance. In the extreme case of total spallation of all available $^{12}$C nuclei, the Li abundance of a companion star would be enhanced up to A(Li)$\sim$6.8 for Solar [C/H] abundance (see equation 3 in \citealt{Boyd91}). A more conservative case can be estimated if the mass-loss rate determines the exposure time of the atoms to the gamma-ray flux (i.e. considering there is no convective mixing). We assume a canonical evaporation/mass-loss rate of $\sim 10^{13}$ g\,s$^{-1}$, a Roche lobe filling secondary star radius of $\sim$0.4\Rsun\ determined from the system parameters of \target\ and a 0.3 percent  efficiency in converting spin-down luminosity into 20\,MeV photons, estimated from the shocked pulsar wind model proposed to explain the UV to GeV gamma-ray spectral energy distribution of \target\ \citep{Takata14}. This results in A(Li)$\sim$4.8, which can be regarded as an upper limit as further inclusion of the effects of a convective layer would diminish the final Li abundance.

Photospallation of heavier elements than $^{12}$C is expected to be less significant than the former, as their energy-weighted cross-sections are $\sim$ 80 times smaller \citep{Boyd89}. If photospallation were the primary cause for the observed $^{16}$O deficit in \target\ spectrum, then the much more efficient photospallation of C would end up destroying all C atoms available on the surface of the secondary star (both intrinsic and proceeding from $^{16}$O). The resultant Li production would then also increase accordingly, generating abundances over A(Li) $\gtrsim$ 6.5. This is at odds with the observations, and would in principle disfavour  photospallation as the main culprit for the under-abundant $^{16}$O. We nevertheless note that our knowledge of the photospallation effects on atoms heavier than $^{12}$C is still limited, and therefore we cannot discard further resonances or production channels that would modify the final measured abundances.

In light of the previous scenarios, we conclude that spallation using high-energy gamma-rays or relativistic protons in the pulsar wind are the most likely mechanism to actively create Li at this stage of \target's evolution, which would allow for a steady state abundance to be reached at the observed level despite the depletion via mass-loss in both accretion- and rotation-powered states. The main puzzle to reconcile is the under-abundant $^{16}$O, which perhaps can be attributed to a fine tuning of the CNO fraction during evolution or from other factors connected to the SN explosion or neutron spallation. It is interesting to note that $^{12}$C tends to plummet right around the phase of binary evolution connected to redbacks ($\sim$ 0.2--0.5\Msun), which means that the production of Li via photospallation or protons might only occur during a certain window of the binary evolution and thus, especially in black widow systems, no Li overabundance might be present.

\subsection{Lithium preservation/depletion}
\label{sec:discussion:depletion}

Once Li has been deposited or created in the atmosphere of the secondary star it can be destroyed or preserved by various processes. We observe a high Li abundance in both the pulsar- and disc-states (see Fig.\,\ref{fig:Li_P_D}), which suggests that once Li is created, the destruction time-scale is sufficiently slow that Li is essentially preserved in the secondary star.

Episodes spent in the accretion-powered state act as a Li sink due to the fact that Li-rich matter is pulled away from the surface of the star via Roche lobe overflow at a faster rate than it can be produced via neutron spallation or other mechanisms, keeping it in equilibrium. Mass transfer feeding the disc is an efficient source of Li depletion. The time-scale for Li enhancement/depletion (see Section\,\ref{sec:discussion:neutrons}) only depends on the mass exposed to the spallation projectile bombardment and the mass-loss rate when the latter is the dominant sink source which typically amounts a few hundred years. Furthermore, it should also be noted that in the rotation-powered pulsar-state, the secondary star is observed to loose mass via an outgassing causing radio eclipses at a typical rate \citep[see, e.g.][]{Polzin20} commensurate to the implied mass accretion rate in the disc-state.

Convection overshooting that mixes Li-rich material from the bottom of the convection zone to regions hot enough for it to be destroyed by nuclear reactions should occur on much larger time-scales of $\sim 10^{7-9}$\,yr \citep[see][and references within]{Yi97}. Numerical simulations of an irradiated stellar surface comparable to that of a redback such as \target\  shows that a boundary layer forms under the photosphere and keeps it largely isolated from the large convective layer underneath, therefore limiting the convective mixing that can take place \citep{Zilles20}. Furthermore, rotation plays a key role in limiting the Li depletion process \citep{Maccarone05} such that the Li we observe in \target\ most likely represents the time-averaged value over its recent history.

Indeed, the rather quick destruction Li time-scales via mass-loss ($\sim$ tens of years) means that to explain the observed enhanced Li, it needs to be created continuously otherwise it will vanish quickly compared to the evolution time of MSPs. One corollary from this is that in `older' spider systems the surface CNO abundance might gradually deplete and impede the regeneration of Li, thus making its abundance gradually decline. The details of this, however, would critically depend on the precise distribution of CNO inside the stars.

\section*{CONCLUSIONS}

We use high-resolution optical spectroscopy of the binary millisecond pulsar to determine the chemical abundances of the secondary star in \target\ which  appears to be tightly controlled by a combination of the rate of angular momentum loss and the initial conditions of the binary.

\begin{itemize}

\item
We determine a metallicity of [Fe/H]=0.48 $\pm$ 0.04 which is higher than the Solar value and measure element abundances in the secondary star that are  different compared to the  secondary stars in X-ray binaries and stars in the Solar neighbourhood.  Compared to the Galactic trends of stars in the Solar neighbourhood with similar Fe content the [Si/Fe] abundance is consistent, the [Ca/Fe], [Al/Fe] and [Ni/Fe] abundances are higher and the [O/Fe] abundance appears to be under-abundant. 

\item
We compare the observed element abundances with the model predictions from different supernova  scenarios, where matter that has been processed in the supernova is captured by the secondary star leading to abundance anomalies, and binary stellar evolution models. We find that the spherical and aspherical supernova models with low mass cuts ($<$1.49\Msun) qualitatively agree  with the observations but we cannot reconcile the O and Al observed abundances. The high element abundances derived in \target, in particular  the high content in Fe-peak elements, support the formation of a neutron star in a SN and so argues against accretion induced collapse formation.

\item
We perform simulations with the binary stellar evolution code \textsc{mesa} and find that the O abundance at the surface of the secondary star appears to be tightly controlled by a combination rate of angular momentum loss and the initial conditions of the binary. In the case of \target\, where we observe hydrogen in the optical spectrum, efficient angular momentum loss must be at play. Under these conditions hydrogen is still present, and we find that the  observed underabundant O is not due to the chemical anomalies associated with the CNO cycle. 

\item
We observe Li in both the pulsar- and disc-state spectrum.  Using the uncontaminated (emission line free) pulsar-state spectrum we determine the Li abundance to be A(Li) = 3.66$\pm$0.20, which is higher than the cosmic value and what is observed in young  Population I stars. This provides unambiguous evidence for fresh Li production. The most likely explanation for the enhanced Li is the interaction of high-energy gamma-rays or relativistic protons in the pulsar wind with the CNO nuclei in the secondary star's atmosphere via spallation which can lead to substantial Li enrichment in the secondary star's atmosphere. The rather quick destruction time-scales over a few hundred years via mass-loss implies that the observed enhanced Li, requires continuous production in the secondary star rather than enrichment through a past event such as the supernova explosion. As most spider pulsars possess energetic winds and are prolific gamma-ray emitters, we would expect higher-than usual Li abundances to be found in other systems as well, though a gradual decay is possible as the CNO mass fraction near the surface might evolve.

\end{itemize}

\section*{ACKNOWLEDGEMENTS}

TS acknowledges acknowledges financial support from the Spanish Ministry of Science and Innovation (MICINN) project PID2020-114822GB-I00.
JIGH acknowledges financial support from the Spanish Ministry of Science and Innovation (MICINN) project PID2020-117493GB-I00, and also from the Spanish MICINN under 2013 Ram\'on y Cajal program RYC-2013-14875.
RPB and MRK acknowledge support of the European Research Council, under the European Union's Horizon 2020 research and innovation program (grant agreement No. 715051; Spiders).
MRK. acknowledges funding from Irish Research Council in the form of a Government of Ireland Postdoctoral Fellowship (GOIPD/2021/670: Invisible Monsters).
DMS  acknowledges the Fondo Europeo de Desarrollo Regional (FEDER) and the Canary Islands government for the financial support received in the form of a grant with number PROID2020010104. 
ML acknowledges funding from the European Research Council (ERC) under the European Union’s Horizon 2020 research and innovation programme (grant agreement No. 101002352).
This work has made use of the VALD database, operated at Uppsala University, the Institute of Astronomy RAS in Moscow, and the University of Vienna.
This paper makes use of data obtained from the Isaac Newton Group Archive which is maintained as part of the CASU Astronomical Data Centre at the Institute of Astronomy, Cambridge. 
Based on data obtained from the ESO Science Archive Facility.
We gratefully acknowledge the use the \textsc{molly} software package developed by Tom Marsh and the \textsc{python} packages: \textsc{matplotlib} \citep{Hunter07}, \textsc{numpy} \citep{vanderWalt11} and \textsc{emcee} \citep{Foreman13},

\medskip

\noindent
{\it Facilities:} WHT (ISIS), VLT (X-SHOOTER)

\section*{Data availability}

The WHT data are publicly available at the ING Archive (http://casu.ast.cam.ac.uk/casuadc/ingarch/). The VLT data are publicly available at the ESO Archive (http://archive.eso.org/). 

\bibliographystyle{mnras}
\bibliography{paper}

\bsp	
\label{lastpage}
\end{document}